\documentclass[12pt, draftclsnofoot, onecolumn]{IEEEtran}
\usepackage{cite,graphicx,amsmath,amssymb}
\usepackage{subfigure}
\usepackage{citesort}
\usepackage{fancyhdr}
\usepackage{mdwmath}
\usepackage{mdwtab}
\usepackage{balance}
\usepackage{xcolor}
\usepackage{bm}

\usepackage{algorithm}
\usepackage{algorithmic}

\usepackage{multirow}
\usepackage{flafter}

\usepackage{multicol}
\usepackage{amsthm}
\usepackage{caption}
\usepackage{graphicx}

\newtheorem{remark}{Remark}

\newtheorem{theorem}{Theorem}

\newtheorem{proposition}{Proposition}
\newtheorem{myDef}{Definition}

\begin{document}


\title{Cache-Aided NOMA Mobile Edge Computing: A Reinforcement Learning Approach}
\author{Zhong\ Yang,~\IEEEmembership{Student Member,~IEEE,}
Yuanwei\ Liu,~\IEEEmembership{Senior Member,~IEEE,}\\
Yue\ Chen,~\IEEEmembership{Senior Member,~IEEE,} and
Naofal\ Al-Dhahir,~\IEEEmembership{Fellow,~IEEE,}

\thanks{ Part of this paper has been presented in IEEE International Communication Conference (ICC) 2019~\cite{zhong2019ICC}.}
\thanks{ Z. Yang, Y. Liu and Y. Chen are with the School of Electronic Engineering and Computer Science, Queen Mary University of London, London E1 4NS, UK. (email:\{zhong.yang, yuanwei.liu, yue.chen\}@qmul.ac.uk)}
\thanks{ N. Al-Dhahir is with the Department of Electrical and Computer Engineering, University of Texas at Dallas, Richardson, TX 75080. (email:~aldhahir@utdallas.edu )}
}
\maketitle

\begin{abstract}
    A novel non-orthogonal multiple access (NOMA) based cache-aided mobile edge computing (MEC) framework is proposed. For the purpose of efficiently allocating communication and computation resources to users' computation tasks requests, we propose a long-short-term memory (LSTM) network to predict the task popularity. Based on the predicted task popularity, a long-term reward maximization problem is formulated that involves a joint optimization of the task offloading decisions, computation resource allocation, and caching decisions. To tackle this challenging problem, a single-agent Q-learning (SAQ-learning) algorithm is invoked to learn a long-term resource allocation strategy. Furthermore, a Bayesian learning automata (BLA) based multi-agent Q-learning (MAQ-learning) algorithm is proposed for task offloading decisions. More specifically, a BLA based action select scheme is proposed for the agents in MAQ-learning to select the optimal action in every state. We prove that the BLA based action selection scheme is instantaneously self-correcting and the selected action is an optimal solution for each state. Extensive simulation results demonstrate that: 1) The prediction error of the proposed LSTMs based task popularity prediction decreases with increasing learning rate. 2) The proposed framework significantly outperforms the benchmarks like all local computing, all offloading computing and non-cache computing. 3) The proposed BLA based MAQ-learning achieves an improved performance compared to conventional reinforcement learning algorithms.
\end{abstract}


\section{Introduction}
Mobile applications have been growing exponentially in wireless networks due to the explosive growth in smart wireless devices. The success of heterogeneous services and related applications, such as augmented reality (AR), virtual reality (VR), real-time online gaming and high-speed video streaming in wireless networks require unprecedented high access speed and low latency~\cite{chen2018wcm,mengting2019twc}. However, the stringent performance and delay requirements of computationally-intensive and latency-sensitive applications are significantly restricted by limited battery capacity and computation resources of the mobile devices. To address these significant challenges, a new trend is emerging with the function of central networks being increasingly moved towards the network edges (see~\cite{mao2017ICST}, and references therein). The key idea of mobile edge computing (MEC) is to promote abundant computing resources at the edge of networks by integrating MEC servers at wireless access points (APs). The computation tasks requested by mobile users can be offloaded to the APs, which liberates the mobile devices from heavy computation workloads and reduces their energy consumption. However, determining communication and computation resources at networks edge introduces significant challenges (see~\cite{mengting2019twc,mao2017ICST}, and references therein), such as task offloading decisions of mobile users, computing resource allocation of the APs to satisfy large numbers of computationally-intensive and latency-sensitive computing tasks..

{\bf For the transmission aspect:} Non-orthogonal multiple access (NOMA) emerged recently as a key enabling technology for next generation wireless communication, thanks to its high bandwidth efficiency and ultra high connectivity~\cite{Qin2018commag,Yuanwei2017pieee}. The key idea behind NOMA is to ensure that multiple users are served simultaneously within the same given time/frequency resource block (RB), utilizing superposition coding (SC) techniques at the transmitter and successive interference cancellation (SIC) at the receiver~\cite{Ding2017commag,xiang2019cachenoma}. Different from orthogonal multiple access (OMA) in MEC networks, NOMA based MEC enables the mobile users to offload the computation tasks to the MEC servers simultaneously, which significantly reduce the computation latency of the network. Therefore, adopting NOMA in MEC networks better utilizes the capacity of the communication channel for offloading computation tasks, and thus improves the computing performance for multiuser MEC networks.

{\bf For the computation aspect:} Caching is another promising technique due to the strategy of trading spectrum resources with storage resources. The main idea behind caching is to place abundant cache resources at the network edge for storing reusable content. In MEC networks, the task computation results may be requested by other users in the near future~\cite{cui2017lcn,mao2016jsac}. For instance, in real-time online gaming, a rendered game frame would be sent to a bunch of nearby users in the same game. Accordingly, caching reusable task computation results in the MEC server is capable of reducing the duplicated task offloading and computing, therefore, it releases the computation burden and latency of the MEC networks~\cite{you2017twc,2018arXiv180506146C}.

\subsection{Related Works}\label{subsection:Related_Works}

\textit{1) Studies on NOMA MEC networks:} In contrast to the conventional OMA based MEC, the NOMA based MEC enables simultaneous task offloading, thus reducing energy consumption and avoiding delay. This motivated numerous researchers to dedicate substantial research contributions to NOMA MEC networks~\cite{Ding2019tcom,Wang2019tcom,Kiani2018iot,Pan2019coml,Ding2018SPL,song2018comletter}. Various asymptotic studies are carried out in~\cite{Ding2019tcom}, revealing that the impact of channel conditions and transmit powers on NOMA MEC are different from conventional NOMA scenarios. For one AP scenario, the authors in~\cite{Wang2019tcom} jointly optimized the communication and computation resource allocation as well as the BS's successive interference cancelation (SIC) decoding order, to minimize the total energy consumption of the users. In~\cite{Kiani2018iot}, the energy consumption of the users is minimized via optimizing the user clustering, computing and communication resource allocation, and transmit powers. Different from~\cite{Wang2019tcom,Kiani2018iot}, which only consider the uplink task transmission, the authors exploited NOMA for both task uploading and result downloading in~\cite{Pan2019coml}. Different from~\cite{Wang2019tcom,Kiani2018iot,Pan2019coml}, which minimize the energy consumption, the authors in~\cite{Ding2018SPL} minimized the offloading delay in a two-user NOMA MEC network. For the multiple APs scenario, a joint radio and computational resource allocation problem is investigated in~\cite{song2018comletter}, aiming at minimizing the energy consumption of all users under the task execution latency in heterogeneous networks. The above works on NOMA MEC analyze the static NOMA MEC networks using stochastic geometry approaches. However, for dynamic NOMA MEC networks, the uncertainty and unavailability of prior information makes conventional approaches difficult or even impossible.

\textit{2) Studies on NOMA Caching networks:} The flexibility of NOMA makes it easy to integrate with other emerging wireless technologies like caching, to enable spectral and energy efficient transmission~\cite{Mojtaba2019nomacahce}. In~\cite{Ding2018tcomnomacahe}, the NOMA strategy is utilized in caching networks, for pushing more contents to the server or simultaneously pushing the contents to the server and users. In~\cite{Zhao2018letter}, the cache-aided NOMA scheme is proposed to improve the coverage performance of NOMA communications. The authors in~\cite{xiang2019cachenoma,Xiang2018nomacache} proposed a cache-aided NOMA scheme, to exploit cached data for interference cancellation in NOMA transmission. In~\cite{Armando2018arxiv}, cache-aided NOMA is proposed to reduce the outage probability when a user possesses a cache of information requested by a user with a stronger channel. Different from~\cite{Mojtaba2019nomacahce,Ding2018tcomnomacahe,Zhao2018letter,xiang2019cachenoma,Xiang2018nomacache,Armando2018arxiv}, which adopt superposition coding (SC) at the transmitters, the authors in~\cite{Fu2019modelselection} utilize index coding (IC) in cache-aided NOMA networks for reducing the transmission power.

\textit{3) Studies on Caching MEC networks:} Wireless caching is typically employed in MEC networks, for supporting multimedia contents in networks edge to reduce the computing overhead and latency~\cite{xiao2018wcom}. The motivation for integrating caching and computing for next generation wireless networks is to facilitate massive content delivery and satiate the requirements of network conditions and mobile devices. Both a cache-assisted computing mechanism and a computing-assisted caching mechanism are proposed in~\cite{Zhou2018wcom}.

\subsection{Motivation}\label{subsection:motivation}

While the aforementioned research contributions have laid a solid foundation for caching, NOMA and MEC networks, the investigations on the applications of integrating caching and NOMA in MEC networks are still in their infancy. It is worth pointing out that, in a multi-users MEC network, where there are several users requesting services from one AP, the major challenge is computing model selection (i.e., local computing or MEC computing) and computation resource allocation. Due to the combinatorial nature of computing mode selection, the task offloading and resource allocation problem is generally formulated as a mixed integer programming (MIP) problem. To tackle this MIP problem, branch-and-bound algorithms~\cite{Narendra1977Tcom} and dynamic programming~\cite{bertsekas2005dynamic} are adopted for the globally optimal solution. Though the conventional approaches make solid contributions to static optimization of task offloading and resource allocation, the mobile users nowadays request dynamic task computing, which is non-trivial for conventional approaches. Furthermore, designing the apriori resource allocation scheme in a long-term manner is nontrivial or even impossible for conventional approaches.

With the development of reinforcement learning (RL) and the high computing speed of new workstations, the investigations on the applications of RL algorithms in wireless networks are growing rapidly~\cite{Qin2019wcom}. RL is a promising approach to find an efficient long-term resource allocation solution in an intelligent manner. In cache-aided NOMA MEC networks, the task popularity prediction is the key foundation to efficiently serve mobile users' dynamic requests.

\textit{1) For the prediction problem:} Long short-term memory (LSTM) networks are utilized in~\cite{Bayati2018letter} to predict the traffic patterns at different timescales. In~\cite{Pichotta2016aaai}, LSTMs are adopted to predict reference information in statistical script. A new spatiotemporal LSTM model is utilized in~\cite{Wang2017nips} to generate future images by learning from the historical frames for spatiotemporal sequences.

\textit{2) For the resource allocation problem:} A model-free deep reinforcement learning (DRL) approach is proposed in~\cite{Bingqian2019aaai} to efficiently allocate resources upon users' requests and price the resource usage. In~\cite{Ghoorchian2019arxiv}, the server selection problem is formulated as a budget-constrained multi-armed bandit (MAB) problem, where each agent is given a reward and cost. A Bayesian reinforcement learning (BRL) algorithm is proposed in~\cite{Asheralieva2017} for a distributed resource sharing problem in a heterogeneous network, which shows a superior performance compared with other resource allocation schemes. It is worth pointing out that the characteristics of cache-aided NOMA MEC networks make it challenging to apply the RL algorithms, because the number of states increase exponentially with the number of users and tasks, which we have demonstrated in our previous work~\cite{zhong2019arxiv}. In this paper, our goal is to answer the following key questions:

\begin{enumerate}

  \item[$\bullet$] $\bf Question~1:$ Do cache-aided NOMA MEC networks significantly outperform non-cached NOMA MEC networks?

  \item[$\bullet$] $\bf Question~2:$ Do NOMA enhanced MEC networks bring energy reduction compared with conventional OMA based networks?

  \item[$\bullet$] $\bf Question~3:$ Do BLA based MAQ-learning algorithm achieve better performance than conventional RL algorithms?

\end{enumerate}

\subsection{Contributions and Organization}

Being motivated to answer the above questions, in this paper, we design a cache-aided NOMA MEC framework, in which cache and computation resources are integrated for computationally-intensive and latency-sensitive tasks. In our proposed framework, we formulate the considered problem as a long-term reward maximization problem that entails joint optimization of task offloading decisions, computation resource allocation and caching decisions. Instead of solving a sophisticated joint optimization problem, the proposed algorithms are capable of learning from the past experience and automatically improve the allocation policy. In this paper, we propose a Bayesian learning automata (BLA) based multi-agent Q-learning (MAQ-learning) algorithm to improve the performance. Furthermore, we utilize the LSTM to predict the task popularity, while in~\cite{zhong2019ICC}, the task popularity is assumed to follow a Zipf distribution. Our main contributions are summarized as follows

\begin{enumerate}
  \item[$\bullet$] We propose a cache-aided NOMA MEC framework to reduce energy consumption. We establish the correlation between task popularity and cache-aided NOMA MEC, by modeling the cache-aided NOMA MEC with task popularity. We formulate a long-term energy minimization problem that jointly optimizes task offloading, computation and cache resource allocation.

  \item[$\bullet$] We propose a LSTM algorithm for task popularity prediction. The task popularity prediction is formulated as a time series prediction problem. We derive the weight value update expressions of LSTM based on a real-time recurrent learning (RTRL) algorithm, due to its low complexity in our proposed task popularity prediction problem.

  \item[$\bullet$] We formulate the total energy consumption as the reward function in the proposed RL algorithms. We prove that the Bayesian learning automata (BLA) based action selection scheme is instantaneously self-correcting and the selected action is an optimal solution for each state.

  \item[$\bullet$] We demonstrate that the proposed cache-aided NOMA MEC framework outperforms the other benchmark schemes such as non-cache MEC networks. The proposed algorithms achieve a performance improvement compared to MAQ-learning algorithms.

\end{enumerate}

The rest of this paper is organized as follows. In Section \ref{section:SystemModel}, the system model for the proposed cache-aided NOMA MEC framework is presented. In Section \ref{section:RNNforprediction}, the LSTMs based task popularity prediction is investigated. The proposed algorithms for cache-aided NOMA MEC is given in Section \ref{section:pfpa}. Simulation results are presented in Section \ref{section:numeralresult}, before we conclude this work in Section \ref{section:Conclusion}. Table \ref{tablealsum} provides a summary of the notations used in this paper.

\begin{table*}[h]
	\caption{LIST OF NOTATIONS}
	\centering
	\begin{tabular}{|c|c|c|c|}\hline
		Notation&Description&Notation&Description\\\hline
		${\mathcal{N}_u}$ & the mobile users& $N_t$ & the computation tasks  \\\hline
        $B$ & the bandwidth between the AP and users & ${{\sigma ^2}}$& the power of additive noise  \\\hline
        ${C_{{\rm{MEC}}}}$ &  the computing capacities of the AP& ${C_{{\rm{Cache}}}}$& the caching capacities of the AP  \\\hline
        ${{\pi _j}}$ & the size of the input of task $j$ (in bits)& ${\bf{T}}_{{\bf{i}},{\bf{j}}}^{{\bf{offload}}}\left( {\bf{t}} \right)$ & the offloading time for task $j$  \\\hline
        ${{\omega _j}}$ & the computing capability required for this task& ${\bf{E}}_{{\bf{i}},{\bf{j}}}^{{\bf{offload}}}\left( {\bf{t}} \right)$& the transmit energy consumption of offloading \\\hline
        ${{\kappa _j}}$ & Caching capacity& ${\bf{T}}_{{\bf{i}},{\bf{j}}}^{{\bf{loc}}}$ &  local computing time for task $j$ \\\hline
        ${\sigma ^2}$ & the computation result of task $j$ (in bits) & ${\bf{E}}_{{\bf{i}},{\bf{j}}}^{{\bf{loc}}}$ & energy consumption of local computing for task $j$  \\\hline
        ${\bf{X}}\left( {\bf{t}} \right)$ &  the task offloading decision & ${\bf T_{i,j}^{mec}\left( t \right)}$ & MEC computing time for task $j$ \\\hline
        ${\bf{Y}}\left( {\bf{t}} \right)$ &  the computing resource allocation vector & ${\bf{E}}_{{\bf{i}},{\bf{j}}}^{{\bf{mec}}}\left( {\bf{t}} \right)$ & energy consumption of MEC computing for task $j$ \\\hline
        ${\bf{Z}}\left( t \right)$ &  the task computation results caching decision vector & ${\bf {{W_f}}}$ & weights in the forgot gate of the LSTM \\\hline
        $\Pr _i^j$ &  the probability that user $i$ requests for task $j$ & ${\bf {{b_f}}}$ & bias in the forgot gate of the LSTM \\\hline
        ${{h_i}\left( t \right)}$ & the channel gain between user $i$ and the MEC server & ${\bf{W_i}}$ & weights of the sigmoid gate in the input gate\\\hline
        ${\bf {R_i}\left( t \right)}$ &  the achievable transmit rate of user $i$ at time $t$ &${\bf W_C}$ &weights of the tanh gate in the input gate \\\hline
        $\rho_{i}\left( t \right)$ & the transmit power of user $i$& ${\bf{b_i}}$ & bias of the sigmoid gate in the input gate\\\hline
        ${\bf b_C}$ & bias of the tanh gate in the input gate& ${\bf {\mu _t}}$ & the learning rate\\\hline

	\end{tabular}
	\label{tablealsum}
\end{table*}
\section{System Model}\label{section:SystemModel}

\subsection{Network Model}\label{subsection:networkmodel}

\begin{figure} [t!]
 \centering
 \includegraphics[width=3.5in]{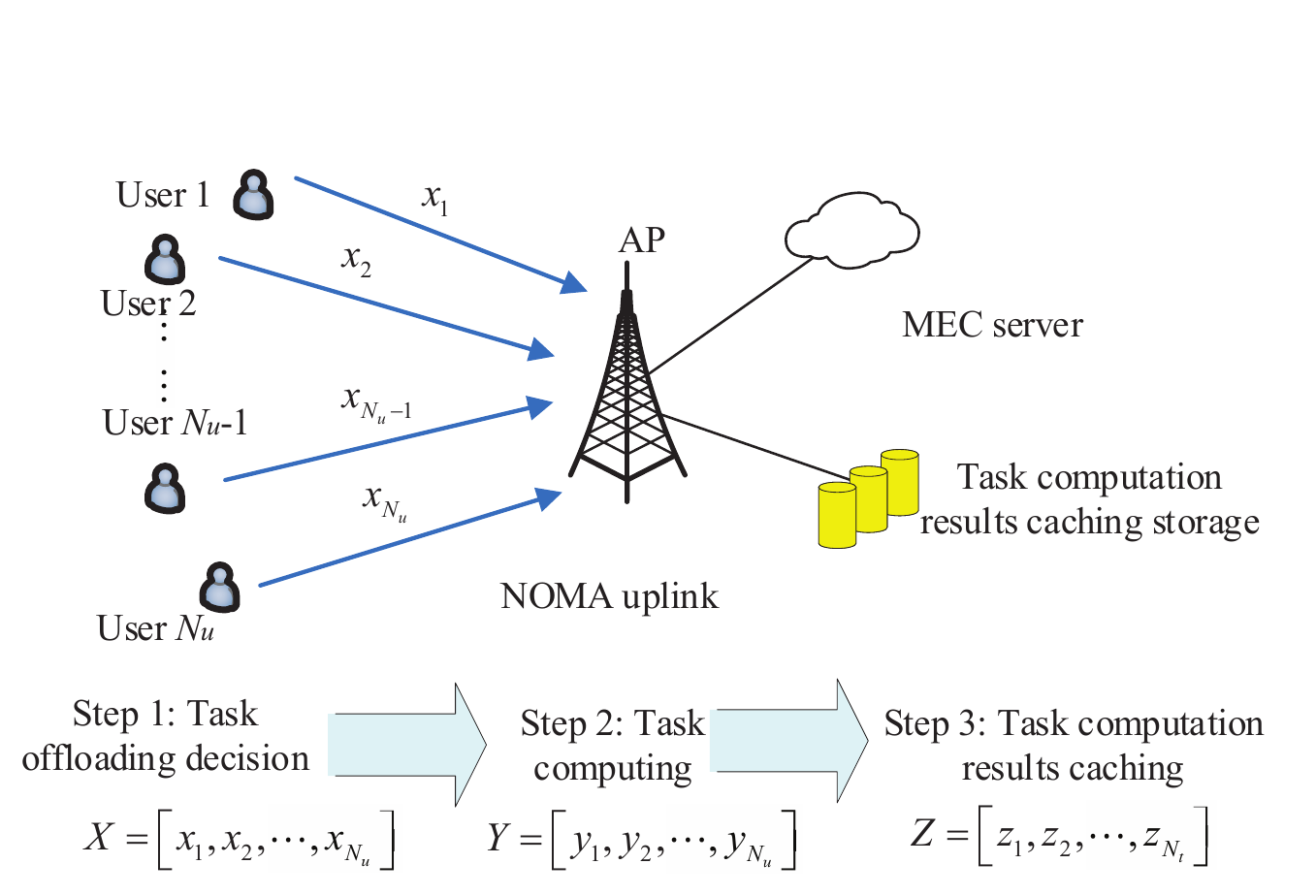}
 \centering
 \caption{An illustration of multi-users cache-aided mobile edge computing networks.}\label{scenarioformec}
\end{figure}

As illustrated in Fig.~\ref{scenarioformec}, we consider a multi-user cache-aided MEC network with a single-antenna AP and $N$ single-antenna users, denoted as ${\mathcal{N}_u} = \left\{ {1, \cdots ,N_u} \right\}$. The bandwidth between the AP and users is denoted by $B$. The AP is associated with an MEC server using optical fiber, whose transmission delay can be ignored. The computing resources are partitioned into a resource pool to provide uniform resources granularity for the users. The AP has strong caching capability, which is capable of caching the computation results to serve other users who request the tasks in the future. The computing and caching capacities of the AP are denoted by ${C_{{\rm{MEC}}}}$ and ${C_{{\rm{Cache}}}}$, respectively. Each user has a limited computation capacity and one intensive computation and sensitive latency computation task. Assume that there are $N_t$ computation tasks in the central network, denoted as ${\mathcal{N}_t} = \left\{ {1, \cdots ,N_t} \right\}$. Each task $j \in {\cal N}_t$ is characterized by three parameters $\left\{ {{\pi _j},{\omega _j},{\kappa _j}} \right\}$, which are defined as follows~\cite{you2017twc}:

\begin{itemize}
  \item ${{\pi _j}}$ represents the size of the input of task $j$ (in bits);
  \item ${{\omega _j}}$ represents the computing capability required for this task which is quantized by the number of CPU cycles;
  \item ${{\kappa _j}}$ represents the computation result of task $j$ (in bits).
\end{itemize}

The task offloading decision is denoted as ${\bf X\left( t \right) }= \left[ {{x_1}\left( t \right),{x_2}\left( t \right), \cdots ,{x_{{N_u}}}\left( t \right)} \right]$, where ${x_i}\left( t \right) \in \left\{ {0,1} \right\}$, ${x_i}\left( t \right) = 0$ means that the task of user $i$ is offloaded to the MEC server for computing, while ${x_i}\left( t \right) = 1$ means that the task is computed locally. The computing resource allocation vector is represented as ${\bf Y\left( t \right)} = \left[ {{y_1}\left( t \right),{y_2}\left( t \right), \cdots ,{y_{{N_u}}}\left( t \right)} \right]$, where ${y_i}\left( t \right) \in \left[ {0,1} \right]$ denotes that the computing resource proportion allocated to user $i$ from the AP. The task computation results caching decision vector is denoted as ${\bf Z}\left( t \right) = \left[ {{z_1}\left( t \right),{z_2}\left( t \right), \cdots ,{z_{{N_t}}}\left( t \right)} \right]$, where ${z_j}\left( t \right) \in \left\{ {0,1} \right\}$, ${z_j}\left( t \right) = 1$ means that the computation result of task $j$ is cached, while ${z_j}\left( t \right) = 0$ means that the result is not cached. If the computation result is cached in the AP, the AP multicasts it to all users, thus reducing the offloading and computing workloads. The probability that user $i$ requests for task $j$ is denoted as $\Pr _i^j \in \left[ {0,1} \right]$~\cite{cui2017lcn}.

\subsection{Communication Model}\label{subsection:communicationmodel}

In our NOMA transmission model, multiple users transmit their own uplink signals to the AP in the same RB. Therefore, the intercellular interference of the users in the same group is identical, the decoding order in one RB is determined only by the channel gains of users. Suppose that there are $N_{up}$ users who choose to upload their computation tasks, represented as ${{\cal N}_{up}} = \left\{ {1, \cdots ,{N_{up}}} \right\}$, where ${N_{up}} = \sum\nolimits_{i = 1}^{N_u} {{x_i}} $. Without loss of generality, assuming the users are ordered as follows:

\begin{equation}\label{users_order}
{\left| {{h_1}\left( t \right)} \right|^2} \ge {\left| {{h_2}\left( t \right)} \right|^2} \ge  \cdots  \ge {\left| {{h_{{N_{up}}}}\left( t \right)} \right|^2},
\end{equation}
where ${\left| {{h_i}\left( t \right)} \right|^2},i \in \left[ {1,{N_{up}}} \right]$ represents the channel gain between user $i$ and the MEC server.

In our NOMA scenario, the user with higher channel gain is decoded first, while the signal of lower channel gain user is considered as interference. Consider user $i \in \left[ {1,{N_{up}}} \right]$ who chooses to upload the computation task at time $t$, then the achievable transmit rate ${\bf {R_i}\left( t \right)}$ (in bits/s) is given by
\begin{equation}\label{rate_of_userm}
{{\bf{R}}_{\bf{i}}}\left( {\bf{t}} \right) = B{\log _2}\left( {1 + \frac{{{\rho _i}\left( t \right){{\left| {{h_i}\left( t \right)} \right|}^2}}}{{\sum\limits_{l = i + 1}^{{N_{up}}} {{\rho _l}\left( t \right){{\left| {{h_l}\left( t \right)} \right|}^2}}  + {\sigma ^2}}}} \right),
\end{equation}
where $\rho_{i}\left( t \right)$ denotes the transmit power of user $i$, and ${{\sigma ^2}}$ represents the power of additive noise. We assume that the users are randomly distributed in a square area. Accordingly, the offloading time for task $j$ with input size ${\pi _j}$ at time $t$ is
\begin{equation}\label{timeoffload}
{\bf T_{i,j}^{offload}\left( t \right)} = \frac{{{\pi _j}}}{\bf{{R_i}\left( t \right)}}.
\end{equation}

In addition, the transmit energy consumption of offloading at time $t$ is given by
\begin{equation}\label{energy_offload}
{\bf{E}}_{{\bf{i}},{\bf{j}}}^{{\bf{offload}}}\left( {\bf{t}} \right) = {\rho _i}\frac{{{\pi _j}}}{{{{\bf{R}}_{\bf{i}}}\left( {\bf{t}} \right)}}.
\end{equation}

\subsection{Computation Model}\label{subsection:computationmodel}

For the computation model, the task $j$ that user $i$ requests can be computed locally on the mobile device or offloaded to the MEC server for computing. Next, we consider the computation overhead in terms of processing time and energy consumption for both local computing and MEC computing.

\subsubsection{Local Computing}\label{subsubsection:loccom}

For user $i$, the local computing capability (i.e. CPU cycles per second) is denoted by $\omega _i^{loc}$, and $P_i^{loc}$ denotes the energy consumption per second for local computing at user $i$. If task $j$ is computed locally, the computing time ${\bf{T}}_{{\bf{i}},{\bf{j}}}^{{\bf{loc}}}$ for task $j$ with computational requirement ${{\omega _j}}$ is
\begin{equation}\label{localcomputtime}
{\bf{T}}_{{\bf{i}},{\bf{j}}}^{{\bf{loc}}} = \frac{{{\omega _j}}}{{\omega _i^{loc}}},
\end{equation}

Hereinafter, the energy consumption of local computing ${\bf{E}}_{{\bf{i}},{\bf{j}}}^{{\bf{loc}}}$ of task $j$ is given by
\begin{equation}\label{localcomputenergy}
{\bf{E}}_{{\bf{i}},{\bf{j}}}^{{\bf{loc}}} = P_i^{loc}\frac{{{\omega _j}}}{{\omega _i^{loc}}},
\end{equation}

\subsubsection{MEC Computing}\label{subsubsection:meccom}

\begin{figure*}[!t]
    \normalsize
    \begin{align}\label{sumenenercon}
    {{\bf{E}}_{\bf{i}}}\left( {t,{x_i}\left( t \right),{y_i}\left( t \right),{z_j}\left( t \right)} \right) = \left( {{\rm{Pr}}_i^j\left( {1 - {z_j}(t)} \right)\left( {{x_i}(t){\bf{E}}_{{\bf{i}},{\bf{j}}}^{{\bf{loc}}} + \left( {1 - {x_i}(t)} \right){\bf{E}}_{{\bf{i}},{\bf{j}}}^{{\bf{offload}}}\left( {\bf{t}} \right) + \left( {1 - {y_i}(t)} \right){\bf{E}}_{{\bf{i}},{\bf{j}}}^{{\bf{mec}}}\left( {\bf{t}} \right)} \right)} \right).
    \end{align}
    \hrulefill \vspace*{0pt}
    \end{figure*}

Let ${y_i}\left( t \right) \in \left[ {0,1} \right]$ denote the proportion of the computing resources that the AP allocated to user $i$. The computing time ${\bf T_{i,j}^{mec}\left( t \right)}$ for task $j$ at time $t$ is
\begin{equation}\label{time_mec}
{\bf{T}}_{{\bf{i}},{\bf{j}}}^{{\bf{mec}}}\left( {\bf{t}} \right) = \frac{{{\omega _j}}}{{{y_i}\left( t \right){C_{MEC}}}}.
\end{equation}

The energy consumption per second for MEC server is denoted as ${P^{mec}}$, thus the energy consumption ${\bf E_{i,j}^{mec}\left( t \right)}$ of task $j$ computed in the AP is given by
\begin{equation}\label{ecl}
{\bf E_{i,j}^{mec}\left( t \right)} = {P^{mec}}\frac{{{\omega _j}}}{{{y_i}\left( t \right){C_{MEC}}}}.
\end{equation}

The computing time for local-execution computing is ${\bf{T}}_{{\bf{i}},{\bf{j}}}^{{\bf{loc}}}$. On the other hand, the computing time ${\bf{T}}_{{\bf{i}},{\bf{j}}}^o\left( {\bf{t}} \right)$ for MEC computing contains two parts ${\bf{T}}_{{\bf{i}},{\bf{j}}}^o\left( {\bf{t}} \right) = {\bf{T}}_{{\bf{i}},{\bf{j}}}^{{\bf{offload}}}\left( {\bf{t}} \right) + {\bf{T}}_{{\bf{i}},{\bf{j}}}^{{\bf{mec}}}\left( {\bf{t}} \right)$. Since the size of the task computation result is smaller than the input, and the downloading data rate is higher than that of the uplink, we neglect the delay and energy consumption associated with results downloading (same assumptions made in~\cite{Wang2018twc,You2016JSAC}). The caching constraint is formulated as $\sum\nolimits_{j = 1}^{N_t} {{z_j}}  = {C_{cache}}$. Also, the computing resources allocating constraint is formulated as $\sum\nolimits_{i = 1}^{{N_u}} {{y_i}}  = 1$. For task $j$ in time $t$, the MEC computing time requirement should satisfy:
\begin{equation}\label{c6}
{\bf{T}}_{{\bf{i}},{\bf{j}}}^{{\bf{offload}}}\left( {\bf{t}} \right) + {\bf{T}}_{{\bf{i}},{\bf{j}}}^{{\bf{mec}}}\left( {\bf{t}} \right) \le T.
\end{equation}
where $T$ denotes the time constraint of task computing.

\begin{figure} [t!]
 \centering
 \includegraphics[width=4.5in]{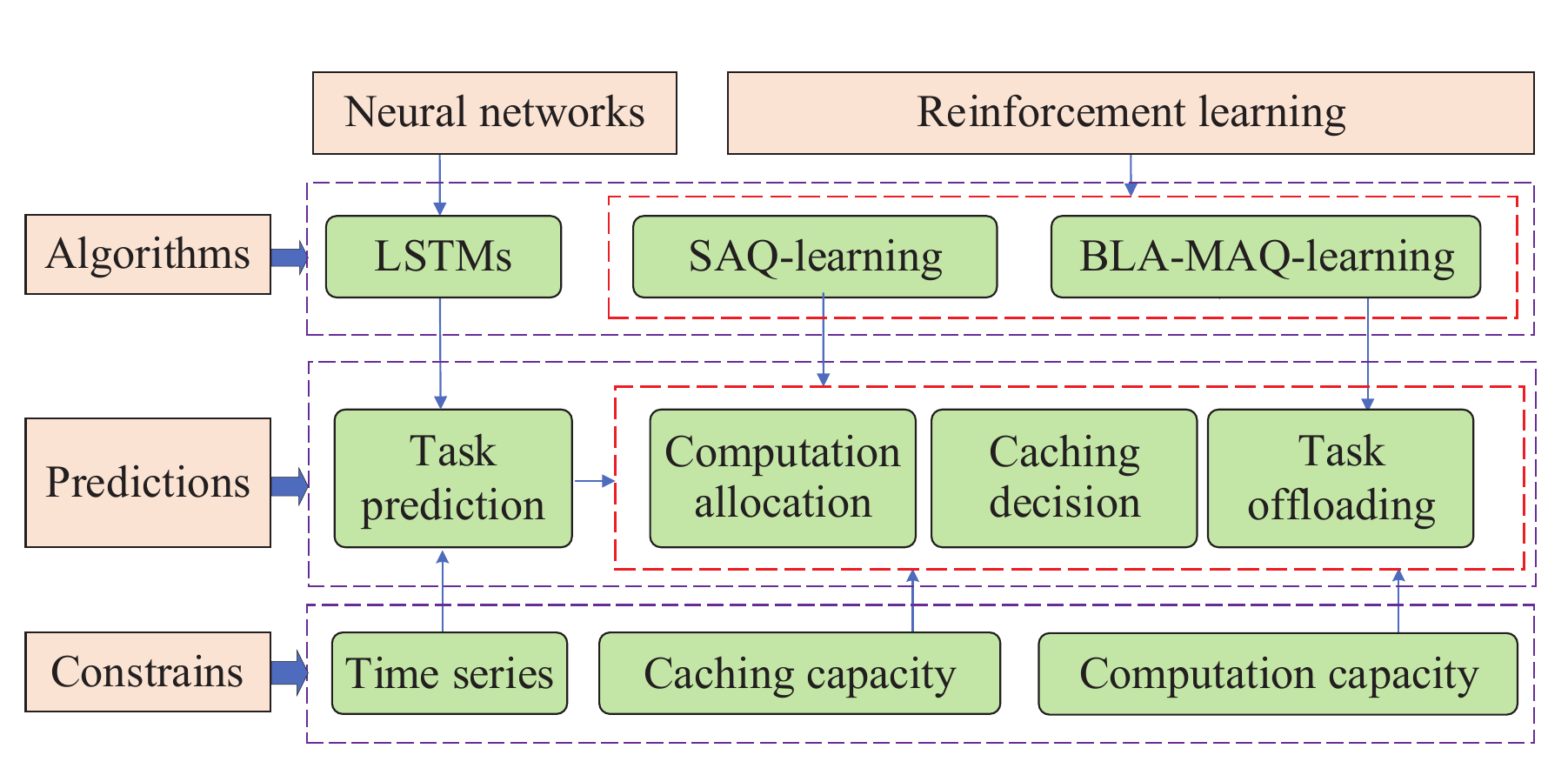}
 \centering
 \caption{An illustration of cache-aided MEC networks.}\label{problems}
\end{figure}

The structure of the cache-aided NOME MEC networks is depicted in Fig.~\ref{problems}. Firstly, we predict the task popularity utilizing the LSTMs. Then, based on the predicted task popularity, we formulate a long-term resource allocation problem, which is solved utilizing a single-agent Q-learning (SAQ-learning) algorithm. Finally, we proposed a BLA based MAQ-learning algorithm for the long-term task offloading problem.
\vspace{-0.3cm}
\section{LSTMs for Task Popularity Prediction}\label{section:RNNforprediction}

\begin{figure} [t!]
 \centering
 \includegraphics[width=4in]{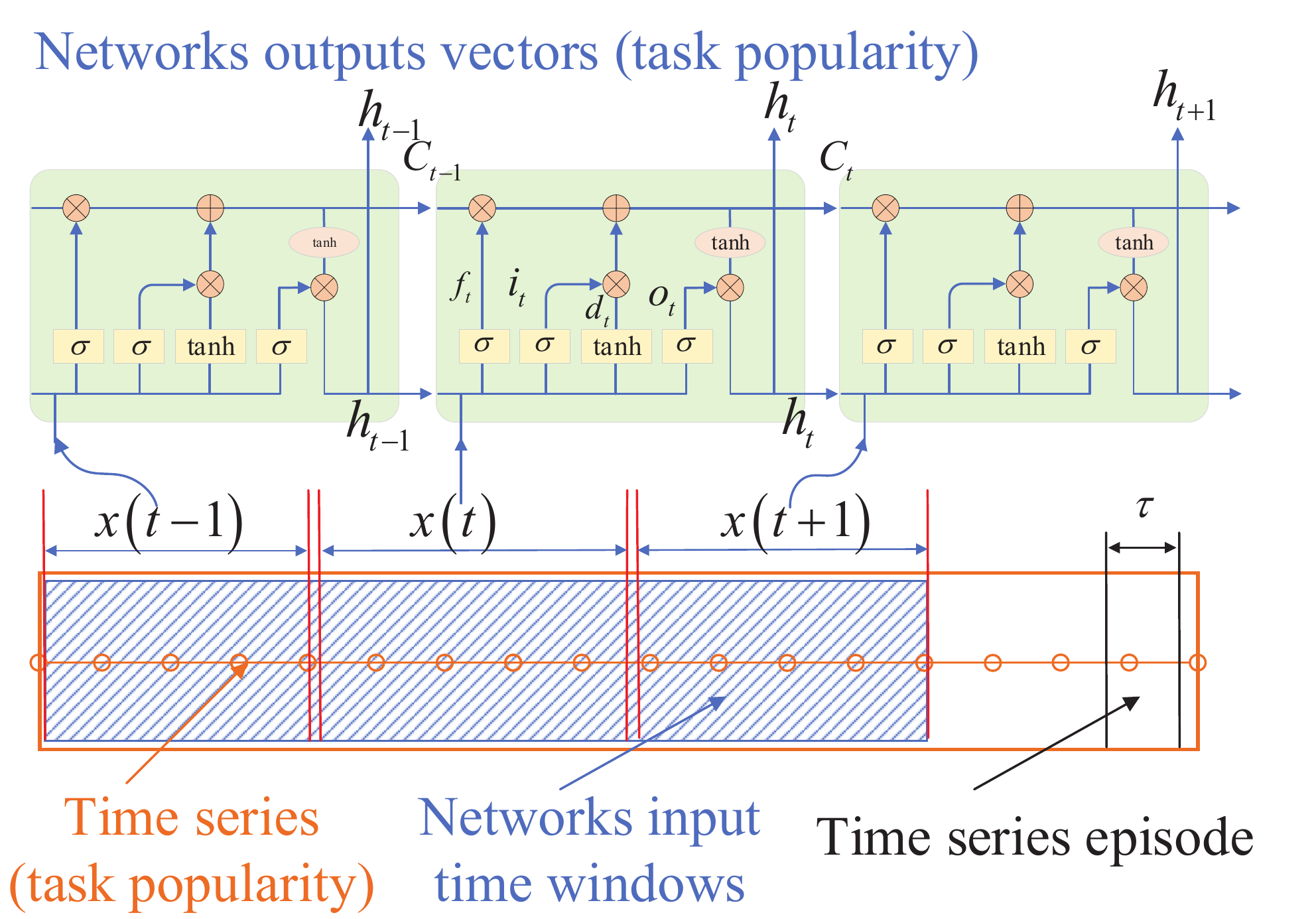}
 \centering
 \caption{Flow chart of LSTMs for task popularity prediction.}\label{flowLSTM}
\end{figure}

Generally, task popularity information can be revealed by direct measurements in real time or by computational estimation of historical data. However, the above techniques suffer from some technical and mercantile issues. Therefore, we predict task popularity from the previously collected data instead of measuring it directly. To predict task popularity, we need to model the map between the historical task popularity and task popularity in the future, which can be used as input sequences and output sequences, respectively. We adopt the widely used LSTMs model to determine the mapping. LSTMs have been widely used for time series prediction like text, speech, audio, etc, to ensure the persistence in learning, by connecting the offline historical data and online data. In our task popularity prediction, we formulate the task popularity as a time series, because it changes in different time slots. In practical situations, we are able to collect the series in a long-term storage. Based on the storage, we are capable of predicting the task popularity in the near future. Since the task popularity evolves over time, we define a dynamic series ${\bf{X_t}} = \left[ {x\left( 1 \right),x\left( 2 \right), \cdots x\left( {{T_p}} \right)} \right]$, where $x\left( t \right) = \left[ {\Pr \left( 1 \right),\Pr \left( 2 \right), \cdots \Pr \left( {{T_s}} \right)} \right]$ denotes a series of task popularity. We assume that time slots are discretized and collect $T_p$ samples of popularity. The input and output of the LSTMs model are presented in Fig.~\ref{flowLSTM}. The goal is to predict ${x_{t + 1}}$ based on our current and past observations $\left[ {x\left( 1 \right),x\left( 2 \right), \cdots x\left( {{T_p}} \right)} \right]$. Assuming that the output of the network is denoted as ${\bf {\widehat x_t}}$, and the observed real task popularity is ${\bf {x_t} }$. Our goal is to minimize ${\bf \sum\nolimits_{t = 1}^T {L\left( {{x_t},{{\widehat x}_t}} \right)}  = \sum\nolimits_{t = 1}^T {{{\left( {{x_t} - {{\widehat x}_t}} \right)}^2}} }$.

The core component of LSTMs to implement the above property is the memory cell state integrated with serval gates to filter different information sequences. This paper follows a similar approach for constructing LSTMs as~\cite{Hochreiter1997}. There are several types of gates in LSTMs, which are described below:

\textbf{Forget gate:} the forget gate decides whether to retain the information by the sigmoid function. The input of the forget gate is ${\bf {x_t}}$ and ${\bf{h_{t - 1}}}$, while the output is a number between $0$ and $1$ for each number in the memory cell state ${\bf C_{t-1}}$.
\begin{equation}\label{forgetgate}
{\bf {f_t} = \sigma \left( {{W_f}\left[ {{h_{t - 1}},{x_t}} \right] + {b_f}} \right)},
\end{equation}
where ${\bf {{x_t}}}$ and ${\bf {{h_{t - 1}}}}$ denote the input of the forget gate. ${\bf {{W_f}}}$ and ${\bf {{b_f}}}$ denotes the weights and bias of the layer, respectively.

\textbf{Input gate:} the input gate decides which information to be committed to memory. The function of input gate is achieved by a sigmoid gate and a tanh gate given below:
\begin{equation}\label{inputgate1}
{\bf {i_t} = \sigma \left( {{W_i}\left[ {{h_{t - 1}},{x_t}} \right] + {b_i}} \right)},
\end{equation}
\begin{equation}\label{inputgate2}
{\bf {d_t} = \tanh \left( {{W_C}\left[ {{h_{t - 1}},{x_t}} \right] + {b_C}} \right)}.
\end{equation}
where ${\bf{W_i}}$ and ${\bf W_C}$ denote the weights of the layers while $\bf{b_i}$ and $\bf{b_C}$ represent the bias of the layers.

\textbf{Output gate:} The output gate decides which information to output. Based on Eq.~(\ref{forgetgate}),~(\ref{inputgate1}),~(\ref{inputgate2}), we obtain the new memory cell state ${\bf C_t}$. The output gate also contains two layers (i.e., a sigmoid layer and a tanh layer).

\begin{algorithm}[h]
\caption{LSTMs based task popularity prediction.}
\label{taskprediction}
\begin{algorithmic}[1]

    \STATE \textbf{Stage One: Back propagation through time (BPTT) training }
\REQUIRE Time interval $T_s$, Training data: task popularity $\left[ {\Pr \left( 1 \right),\Pr \left( 2 \right), \cdots \Pr \left( {{T_p}} \right)} \right]$.

    \STATE Randomly initialize weight matrix ${\bf {W_f},{W_i},{W_C},{W_o}}$ and bias ${\bf {b_f},{b_i},{b_C},{b_o}}$ in the hidden layers.
    \FOR{each episode}
    \STATE Choose input $x\left( t \right)$ from training set;\
    \STATE Forward propagation algorithm: calculate the output of the layers ${\bf {h_t}}$;\
    \STATE Calculate loss function $\sum\nolimits_{{\bf{t}} = {\bf{1}}}^{\bf{T}} {{\bf{L}}\left( {{{\bf{x}}_{\bf{t}}},{{{\bf{\hat x}}}_{\bf{t}}}} \right)} $;\
    \STATE BPTT algorithm: adjust the weights ${\bf {W_f},{W_i},{W_C},{W_o}}$ and bias ${\bf {b_f},{b_i},{b_C},{b_o}}$ according to the loss function;\
    \ENDFOR
    \ENSURE Weights ${\bf {W_f},{W_i},{W_C},{W_o}}$ and bias ${\bf {b_f},{b_i},{b_C},{b_o}}$ of the network.\

    \STATE \textbf{Stage Two: Testing}
\REQUIRE Time interval $T_s$, Testing data: task popularity, Parameters of s: weights ${\bf {W_f},{W_i},{W_C},{W_o}}$ and bias ${\bf {b_f},{b_i},{b_C},{b_o}}$ of the network, number of epochs.

    \FOR{each episode}
    \STATE Choose input ${\bf{X}}$ from testing set;\
    \STATE Forward propagation: calculate the output of the layers ${\bf {h_t}}$;\
    \STATE Calculate loss function $\sum\nolimits_{{\bf{t}} = {\bf{1}}}^{\bf{T}} {{\bf{L}}\left( {{{\bf{x}}_{\bf{t}}},{{{\bf{\hat x}}}_{\bf{t}}}} \right)} $;\
    \ENDFOR
    \ENSURE Task popularity in the next time slot $\left[ {\Pr \left( {{T_p} + 1} \right),\Pr \left( {{T_p} + 2} \right), \cdots \Pr \left( {{T_p} + {T_s}} \right)} \right]$.
\end{algorithmic}
\end{algorithm}

\begin{equation}\label{outputgate1}
{\bf {C_t} = {f_t}{C_{t - 1}} + {i_t}{d_t},}
\end{equation}
\begin{equation}\label{outputgate2}
{\bf {o_t} = \sigma \left( {{W_o}\left[ {{h_{t - 1}},{x_t}} \right] + {b_o}} \right),}
\end{equation}
\begin{equation}\label{outputgate3}
{\bf {h_t} = {o_t} * \tanh \left( {{C_t}} \right).}
\end{equation}
where ${\bf {C_t}}$ is the state of the cell, ${\bf {o_t}}$ is the combination of history output and input of this layer, ${\bf {h_t}}$ is the output of the network, ${\bf W_o}$ and ${\bf b_o}$ denotes the weights and bias.

\begin{remark}\label{remark:LSTM}
The mobile users request different tasks in different time slots, thus the task popularity prediction is a time series prediction problem. The balance ability between historical information and instantaneous information of the proposed LSTMs makes it capable of predicting the task popularity in the cache-aided NOMA MEC networks.
\end{remark}

Given the output of the network ${\bf {h_t}}$, the estimated output is obtained as
\begin{equation}\label{estimatedoutput}
\begin{array}{l}
\bf {\widehat x_t} = {w_t}{h_t}\\
{\kern 1pt} {\kern 1pt} {\kern 1pt} {\kern 1pt} {\kern 1pt} {\kern 1pt} {\kern 1pt} {\kern 1pt} {\kern 1pt} {\kern 1pt} {\kern 1pt} {\kern 1pt} {\kern 1pt}  = {\bf {w_t}{o_t}*\tanh \left( {{C_t}} \right)},
\end{array}
\end{equation}
where ${\bf w_t}$ is the regression coefficient.

To obtain the corresponding parameters above, we introduce the real-time recurrent learning (RTRL) algorithm~\cite{Williams1997}, which enjoys low complexity for our proposed task popularity prediction scenario. For the weight vector $\bf{w_t}$ in Eq.~(\ref{estimatedoutput})
\begin{equation}\label{weightvector1}
\begin{aligned}
{\bf {w_{t + 1}}} & = {\bf {w_t} - {\mu _t}{\nabla _{{w_t}}}l\left( {{x_t},{{\widehat x}_t}} \right)}\\
& = {\bf {w_t} - {\mu _t}{\nabla _{{w_t}}}{\left( {{x_t} - {{\widehat x}_t}} \right)^2}}\\
& = {\bf {w_t} + 2{\mu _t}\left( {{x_t} - {{\widehat x}_t}} \right){o_t}\tanh \left( {{C_t}} \right)},
\end{aligned}
\end{equation}
where ${\bf {\mu _t}}$ represents the learning rate. We set ${\bf {\mu _t} = {\raise0.7ex\hbox{$1$} \!\mathord{\left/
 {\vphantom {1 t}}\right.\kern-\nulldelimiterspace}
\!\lower0.7ex\hbox{$t$}}}$.

\begin{proposition}\label{proposition1}
\emph{For each element $w_C$ in ${\bf {W_C}}$, we update it as follows}
\begin{equation}\label{weightvector2}
{\bf {w_C} = {w_C} + 2{\mu _t}\left( {{x_t} - {{\widehat x}_t}} \right){w_t}\frac{{\partial \left( {{o_t}\tanh \left( {{C_t}} \right)} \right)}}{{\partial \left( {{w_C}} \right)}}}.
\end{equation}
\begin{remark}\label{remark:diversity}
The partial derivative in Eq.~(\ref{weightvector2}) contains two parts: the first is the output ${\bf {{o_t}}}$, and the second part is the memory cell state ${\bf \tanh \left( {{C_t}} \right)}$. We can implement Eq.~(\ref{weightvector2}) if we compute ${\bf \frac{{\partial \left( {{o_t}} \right)}}{{\partial \left( {{w_C}} \right)}}}$ and ${\bf \frac{{\partial \left( {\tanh \left( {{C_t}} \right)} \right)}}{{\partial \left( {{w_C}} \right)}}}$
\end{remark}
\begin{IEEEproof}
See Appendix A~.
\end{IEEEproof}
\end{proposition}

\vspace{-0.3cm}
\section{Resource Allocation Problem Formulation and Proposed Approaches}\label{section:pfpa}

\subsection{Problem Formulation}\label{subsection:pf}

Based on the predicted task popularity we obtained from Section~\ref{section:RNNforprediction}, we formulate the optimization problem for cache-aided NOMA MEC networks as in~(\ref{optimizationproblem}) below, in which, the objective function is the energy consumption given by Eq. (\ref{sumenenercon}). The parameters are the offloading decision vector, computation resources allocation vector and caching vector. The joint task offloading decision and resource allocation problem is formulated as follows:

\begin{subequations}\label{optimizationproblem}
\begin{align}
& \left( {{\bf{P1}}} \right)\mathop {{\rm{min}}}\limits_{X,Y,Z} \sum\limits_{t = 1}^T {\sum\limits_{i = 1}^{{N_u}} {{{\bf{E}}_{\bf{i}}}\left( {t,{x_i}\left( t \right),{y_i}\left( t \right),{z_j}\left( t \right)} \right)} } ,\label{optimization_problem}\\
\mbox{s.t.} \quad
& C_1:\;{x_i}\left( t \right) \in \left\{ {0,1} \right\},\forall i \in \left[ {1,{N_u}} \right],t \in \left[ {1,T} \right],\label{cc1}\\
& C_2:\;{y_i}\left( t \right) \in \left[ {0,1} \right],\forall i \in \left[ {1,{N_u}} \right],t \in \left[ {1,T} \right], \label{cc2}\\
& C_3:\;{z_j}\left( t \right) \in \left\{ {0,1} \right\},\forall j \in \left[ {1,{N_t}} \right],t \in \left[ {1,T} \right],\label{cc3}\\
& C_4:\;\sum\limits_{i = 1}^{{N_u}} {{y_i}} \left( t \right) = 1,\forall t \in \left[ {1,T} \right],\label{cc4}\\
& C_5:\;\sum\limits_{j = 1}^{N_t} {{z_j}\left( t \right)}  \le {C_{{\rm{cache}}}},\forall t \in \left[ {1,T} \right],\label{cc5}\\
& C_6:\;{\bf T_j^o\left( t \right)} \le T,\forall j \in \left[ {1,{N_t}} \right],t \in \left[ {1,T} \right].\label{cc6}
\end{align}
where ${{\bf{E}}_{\bf{i}}}\left( {t,{x_i},{y_j},{z_j}} \right)$ denotes the energy consumption for computing task $j$ given in Eq.~(\ref{sumenenercon}), ${x_i}\left( t \right)$ denotes the offloading decision of user $i$ at time $t$, and ${y_i}\left( t \right)$ denotes the computing speed that the AP allocates to user $i$ at time $t$. The variable ${z_j}\left( t \right)$ denotes the task caching decision of task $j$ at time $t$. Eq.~(\ref{cc4}) guarantees that the task computing resource allocation is valid. While Eq.~(\ref{cc5}) guarantees that the task computation results cached do not exceed the caching capacity of the AP. Finally, Eq.~(\ref{cc6}) guarantees the time constraint.
\end{subequations}

\begin{remark}\label{remark:long-termminimizationprobelm}
The formulated problem is a long-term minimization problem. Consequently, in order to solve it, we need to design a long-term policy, which instantaneously gives the optimal  offloading decision vector, computation resources allocation vector and caching vector at different time slots.
\end{remark}

\subsection{SAQ-learning Solution for Cache-aided NOMA MEC}\label{subsection:mdpcamec}

In this section, a single-agent Q-learning (SAQ-learning) algorithm is proposed for the formulated problem in {\bf Subsection~\ref{subsection:pf}}. We reformulate the cache-aided NOMA MEC problem as a SAQ-learning model, which consists of three elements: states, actions, and rewards. The aim of this section is to devise a resource allocation policy ${\pi ^*}$ that is capable of quickly generating an optimal resource allocation action.

\begin{figure*}[!t]
    \normalsize
    \begin{align}\label{Qvalueupdate}
        {Q^t}\left( {s,a} \right) = \left( {1 - \gamma } \right)\underbrace {{Q^{t - 1}}\left( {s,a} \right)}_{{\rm history{\kern 1pt} {\kern 1pt} Q{\kern 1pt} {\kern 1pt} value}} + \underbrace \gamma _{{\rm learning{\kern 1pt} {\kern 1pt} rate}}\left[ {\underbrace {{r_t}}_{{\rm obtained{\kern 1pt} {\kern 1pt} reward}} + \underbrace \beta _{{\rm discount{\kern 1pt} {\kern 1pt} factor}}\underbrace {\mathop {\max }\limits_{{a_{t + 1}}} {Q^{t - 1}}\left( {{s_{t + 1}},{a_{t + 1}}} \right)}_{{\rm optimal{\kern 1pt} {\kern 1pt} Q{\kern 1pt} {\kern 1pt} value{\kern 1pt} {\kern 1pt} {\kern 1pt} {\kern 1pt} in{\kern 1pt} {\kern 1pt} the{\kern 1pt} {\kern 1pt} next{\kern 1pt} {\kern 1pt} state}}} \right].
    \end{align}
    \hrulefill \vspace*{0pt}
\end{figure*}

\begin{myDef}\label{Def.MDP}
A SAQ-learning model is a tuple $\left\langle {{\cal S},{\cal A},{\cal R}} \right\rangle $. The definitions of the three parameters are given below.
\begin{itemize}
  \item State space ($\mathcal{S}$): A finite set of states of the agent; We assume that all the users are controlled by a central agent, the state of agent is denoted as ${s_t} = \sum\limits_{i = 1}^{{N_u}} {{{\bf{E}}_{\bf{i}}}\left( {t,{x_i}\left( t \right),{y_i}\left( t \right),{z_j}\left( t \right)} \right)} $. The state is the energy consumption of the network in time slot $t$.
  \item Action space ($\mathcal{A}$): A finite set of actions. The action of the central agent is defined as ${a_t} = \left[ {{a_1}\left( t \right),{a_2}\left( t \right),{a_3}\left( t \right)} \right]$, where, the first part ${a_1}\left( t \right) = \left[ {{x_1}\left( t \right), \cdots ,{x_{N_u}}\left( t \right)} \right]$ is the task offloading vector of the users. The second part ${a_2}\left( t \right) = \left[ {{f_1}\left( t \right), \cdots ,{f_{N_u}}\left( t \right)} \right]$ contains the computing speed allocation from the AP to all users. The last part ${a_3}\left( t \right) = \left[ {{c_1}\left( t \right), \cdots ,{c_{N_u}}\left( t \right)} \right]$ represents the cache decision of the AP.
  \item Reward function (${\cal R}$): The reward function is the key to solve $\left( {{\bf{P1}}} \right)$ by utilizing the SAQ-learning model. The objective of SAQ-learning is to maximize the expected  long-term reward, also known as the state-value function given by:

        \begin{equation}\label{state_valuefunction}
        \begin{aligned}
        {V_\pi }(s) &= {\mathbb{E}_\pi }\left[ {\sum\limits_{t = 0}^\infty  {{\gamma ^t}{r_t}\left| {s_0 = {s}} \right.} } \right] \\
        & = {\mathbb{E}_\pi }\left[ {{r_0} + \gamma E\left[ {\gamma {r_1} + {\gamma ^2}{r_2} + ...} \right]\left| {s_0 = {s}} \right.} \right] \\
        & = {\mathbb{E}_\pi }\left[ {r\left( {s'\left| {{s_0},a} \right.} \right) + \gamma {V^\pi }(s')\left| {s_0 = {s}} \right.} \right].
        \end{aligned}
        \end{equation}
        where $\gamma $ represents the learning rate and ${{r_t}}$ denotes the reward that the agents obtain in time slot $t$. In order to minimize the energy consumption in $\left( {{\bf{P1}}} \right)$, we define the reward function as ${r_t} = \sum\nolimits_{i = 1}^{{N_u}} {{{\bf{E}}_{\bf{i}}}\left( {t - 1,{s_{t - 1}}} \right)}  - \sum\nolimits_{i = 1}^{{N_u}} {{{\bf{E}}_{\bf{i}}}\left( {t,{s_t}} \right)}  $.

  \item Action value function (${Q^\pi }(s,a)$): The long-run average reward is the payoff of an infinite path where every state-action pair is assigned a real-valued reward.

    \begin{equation}\label{Vpis2}
    {Q^\pi }(s,a) = {\mathbb{E}_\pi }\left[ {\sum\limits_{i = 0}^\infty  {{\gamma ^i}{r_i}\left| {s = {s_0},a = {a_0}} \right.} } \right].
    \end{equation}

  \item A policy $\pi$ is a distribution over actions given states,

    \begin{equation}\label{policypi}
    \pi \left( {a|s} \right) = P\left[ {{A_t} = a|{S_t} = s} \right],
    \end{equation}
    where a policy fully defines the behaviour of an agent.

  \end{itemize}
\end{myDef}

Therefore, the optimal policy is given by

\begin{equation}\label{pi}
{\pi ^*}\left( s \right) = \mathop {\arg \min }\limits_\alpha  {V_\pi }(s),\forall s \in S,
\end{equation}
which finds an optimal policy (i.e., a set of actions) to maximize the long-run average reward.

\begin{table*}[!htb]
	\caption{Q-table of Q-learning Algorithm}
	\centering
	\begin{tabular}{|c|c|c|c|c|}\hline
		
    & ${a_1}$ & ${a_2}$ & ... & ${a_{N_2}}$  \\\hline

    ${s_1}$ & ${Q^{1,1}}\left( {{s_1},{a_1}} \right)$ & ${Q^{1,2}}\left( {{s_1},{a_2}} \right)$ & ... & ${Q^{1,N_2}}\left( {{s_1},{a_{N_2}}} \right)$  \\\hline

    ${s_2}$ & ${Q^{2,1}}\left( {{s_2},{a_1}} \right)$ & ${Q^{2,2}}\left( {{s_2},{a_2}} \right)$ & ... & ${Q^{2,N_2}}\left( {{s_2},{a_{N_2}}} \right)$  \\\hline

     ... & ... & ... & ... & ...  \\\hline

    $s_{N_1}$ & ${Q^{{N_1},1}}\left( {{s_{{N_1}}},{a_1}} \right)$ & ${Q^{{N_1},2}}\left( {{s_{{N_1}}},{a_2}} \right)$ & ... & ${Q^{{N_1},N_2}}\left( {{s_{{N_1}}},{a_{N_2}}} \right)$  \\\hline

	\end{tabular}

	\label{tableQtable}
\end{table*}

According to \cite{Sutton2016rl}, the optimal state-value function ${V^\pi }\left( s \right)$ can be achieved by solving the Bellman's optimality equation.

\begin{remark}\label{remark:optimalremark}
The optimal state-value function ${V^\pi }\left( s \right)$ satisfies the Bellman's optimality equation, which is given by

\begin{equation}\label{remark1}
{V^\pi }\left( s \right) = \mathop {\min }\limits_{a \in A} \left\{ {\left( {1 - \gamma } \right)p\left( {x,y} \right) + \gamma \sum\limits_{s' \in S} {\Pr \left\{ {s'|s,a} \right\}V\left( {s'} \right)} } \right\}.
\end{equation}
\end{remark}

In order to solve this problem, Bellman's equations and backward induction are used. We develop a SAQ-learning based resource allocation algorithm for cache-aided NOMA MEC system. In SAQ-learning, the Q value ${Q^t}\left( {s,a} \right)$ at time slot $t$ is updated as in Eq.~(\ref{Qvalueupdate}). The SAQ-learning algorithm for cache-aided NOMA MEC includes two cycles: the first cycle in each episode obtains the optimal policy, while the second cycle in each step executes the policy in each step. The central agent has to decide all the actions from the first to the last time slot to minimize the total long-term energy consumption. The details of the SAQ-learning are presented in our previous work~\cite{zhong2019arxiv}, and are omitted here due to space limitations. The size of the Q-table is shown in Table~\ref{tableQtable}, in which ${N_1} = {2^{{N_u}}}{N^{{N_f} }}$, ${N_2} = 2 {{N_u}}  {{N_f} } $, where $N_u$ represent the number of users and, $N_f$ denotes the number of slices the computing resources are sliced into.

\subsection{BLA-MAQ-learning Solution for Task Offloading}\label{section:ADLS}

To reduce the dimension of state space and action space of the agent, which reduce the performance of learning~\cite{zhong2019arxiv}, we propose a distributed RL algorithm, i.e., MAQ-learning algorithm, for task offloading. In MAQ-learning, the users operate cooperatively as distributed agents to decide whether to offload the computation tasks to the server or not.

\begin{myDef}\label{Def.MDP}
A MAQ-learning model is a tuple $\left\langle {{\cal S},{\cal A},{\cal R}} \right\rangle $. The definition of the parameters adopted in MAQ-learning are given below.

\begin{itemize}
  \item State ($\mathcal{S}$): We assume that all the users are set as agents, the state of agent $i$ is denoted as $s_t^i = {{\bf{E}}_{\bf{i}}}\left( t \right)$. The state is the energy consumption of computing of the task that the user requests.
  \item Action ($\mathcal{A}$): The action of the agent $i$ is defined as $a_t^i \in \left[ {0,1} \right]$, where $a_t^i{\rm{  =  1}}$ implies that the task is computed locally while $a_t^i{\rm{  =  0}}$ implies that the task is offloaded to the AP for computing.
  \item Reward function (${\cal R}$): The local reward function of the central agent is defined as $r_t^i = {{\bf{E}}_{\bf{i}}}\left( {t - 1,s_{t - 1}^i} \right) - {{\bf{E}}_{\bf{i}}}\left( {t,s_t^i} \right)$.
  \end{itemize}
\end{myDef}

\begin{figure*}

    \centering

    \subfigure[Bayesian learning automata based action selection scheme.]{
    \begin{minipage}{7cm}\label{blaactionselectionleft}
    \centering
    \includegraphics[width=7cm]{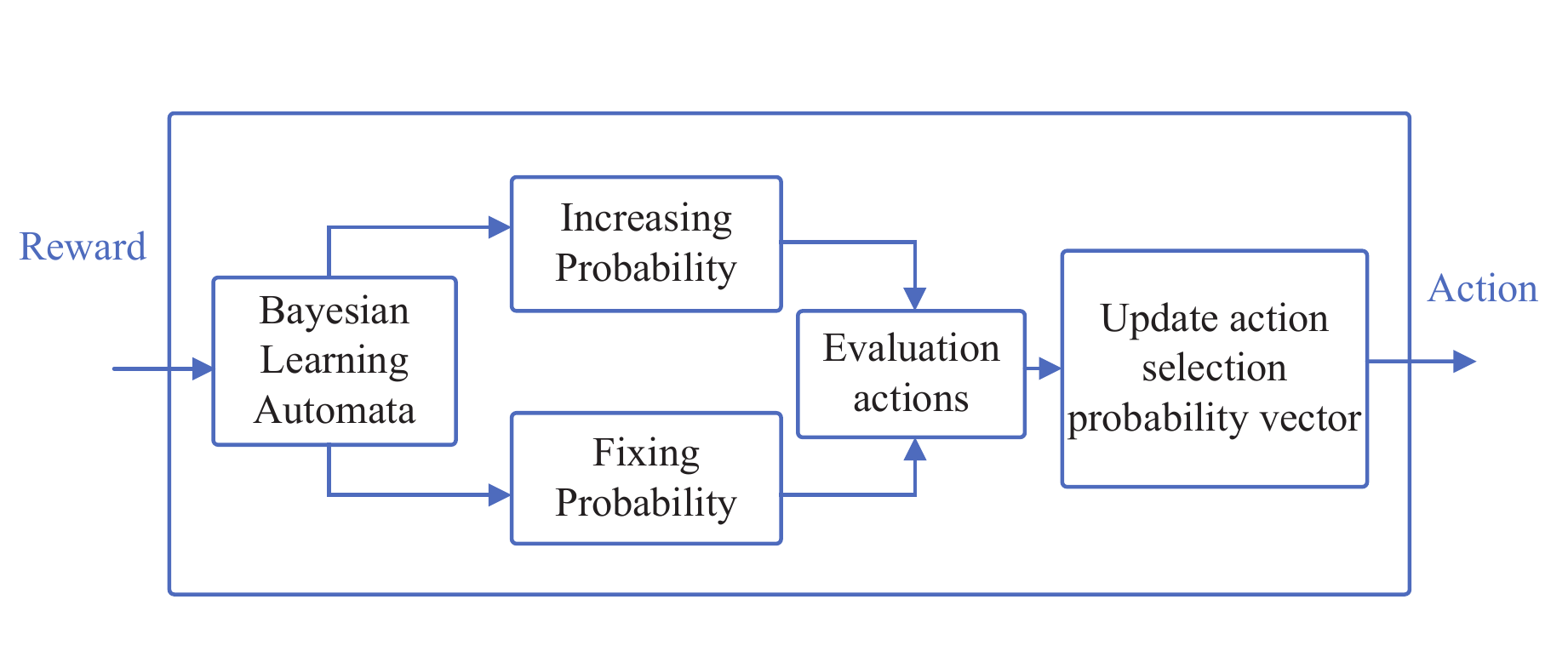}
    \end{minipage}
    }
    \subfigure[BLA-MAQL for cache-aided NOMA-MEC networks.]{
    \begin{minipage}{7cm}
    \centering
    \includegraphics[width=7cm]{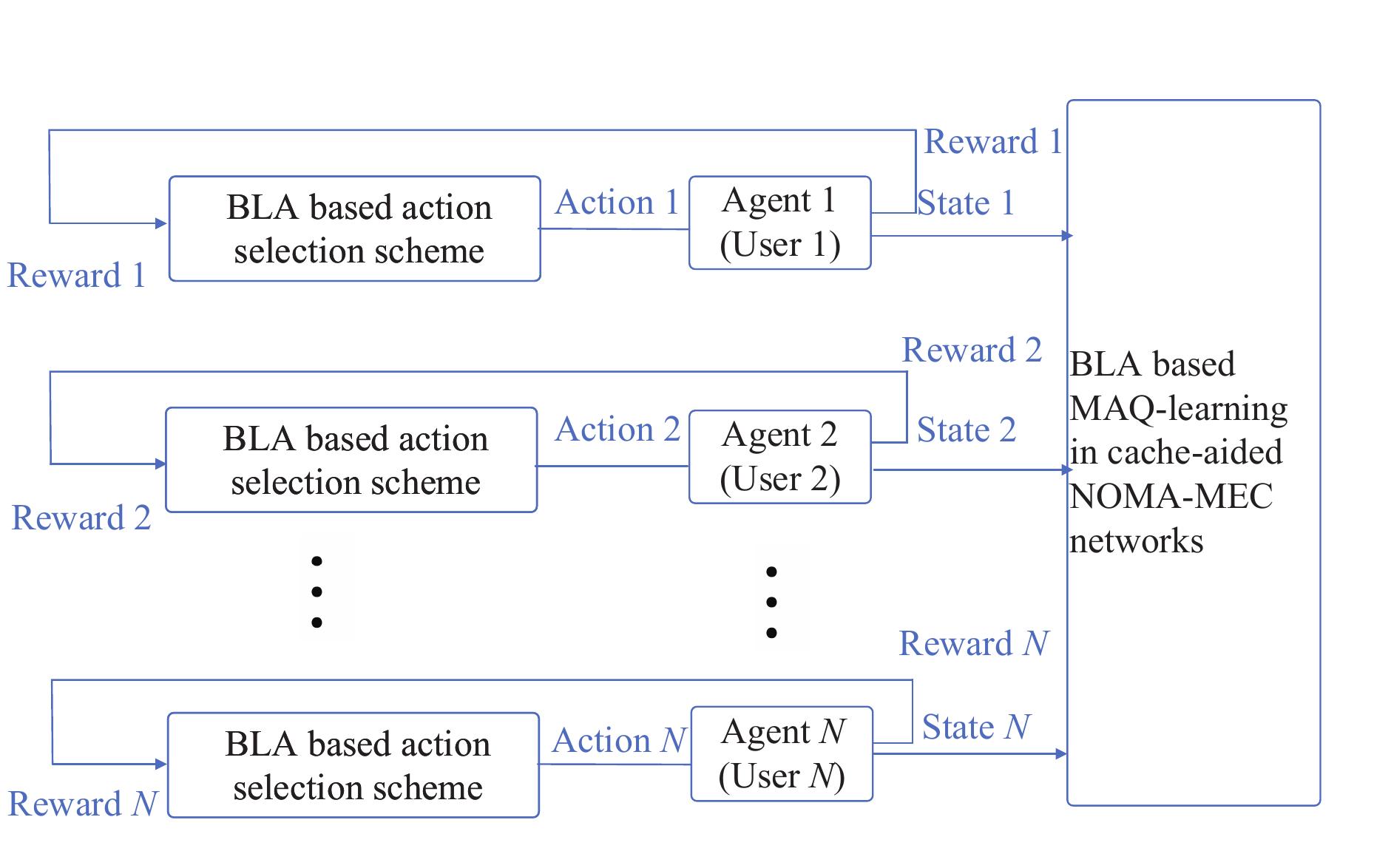}
    \end{minipage}
    }
    \caption{An illustration of Bayesian learning automata based multi-agent Q-learning in cache-aided NOMA-MEC networks.}\label{illustrationQlearningcachingM2cc3}
\end{figure*}

Different from SAQ-learning, in the MAQ-learning case, three cycles are looped over. Apart from the two cycles in SAQ-learning, one more loop for the agents is added, to cooperatively decide the actions, as shown in Fig. \ref{illustrationQlearningcachingM2cc3}. The action selection scheme (i.e., how the action is selected during the learning process) is the core issue in the RL algorithm, which is utilized to balance the exploration and exploitation and avoiding over-fitting. In conventional RL algorithms, $\epsilon$-greedy exploration is adopted. However, the selection mechanism of $\epsilon$-greedy exploration is based on a random mechanism. In order to overcome this drawback, we apply Bayesian learning automata (BLA) for action selection, because BLA is capable of choosing the optimal action for two action case~\cite{Granmo2008}.

A BLA is an adaptive decision-making unit which learns the optimal action from a set of actions offered by the environment it operates in. Essentially, there are two Beta distributions in the BLA based action selection scheme. In our MEC scenario, we assume that the computation task is computed locally as Arm 1, while the task is offloaded to MEC server for computing as Arm 2. The heart of BLA is the beta distribution, whose probability density function is:

\begin{equation}\label{probabilitydensityfunction}
\begin{array}{*{20}{c}}
{f\left( {x;\alpha ,\beta } \right) = \frac{{{x^{\alpha  - 1}}{{\left( {1 - x} \right)}^{\beta  - 1}}}}{{\int_0^1 {{u^{\alpha  - 1}}{{\left( {1 - u} \right)}^{\beta  - 1}}du} }},}&{x \in \left[ {0,1} \right]}
\end{array},
\end{equation}
where $\alpha $ and $\beta $ are the parameters of beta distribution and the corresponding cumulative distribution function is:

\begin{equation}\label{cumulativedistributionfunction}
\begin{array}{*{20}{c}}
{F\left( {x;\alpha ,\beta } \right) = \frac{{\int_0^x {{t^{\alpha  - 1}}{{\left( {1 - t} \right)}^{\beta  - 1}}dt} }}{{\int_0^1 {{u^{\alpha  - 1}}{{\left( {1 - u} \right)}^{\beta  - 1}}du} }},}&{x \in \left[ {0,1} \right]}
\end{array}.
\end{equation}

BLA uses the beta distribution for two purposes. Firstly, the beta distribution is used to provide a Bayesian estimation of the reward probabilities associated with each action of Q-learning. Secondly, a novel feature of BLA is that it uses the beta distribution as the basis for a randomized action selection mechanism. The state of user $i$ at time slot $n$ is $s_i^n = \left( {\alpha _1^n,\beta _1^n,\alpha _2^n,\beta _2^n} \right)$, where $({\alpha _1^n,\beta _1^n})$ determines the parameter of the first Beta distribution, and $({\alpha _2^n,\beta _2^n})$ determines the parameter of the second Beta distribution. The action selection probability of Arm 1 (i.e., local computing) is given in Eq.~(\ref{Probabilityofarm1}). After taking the action, the parameters of the Beta distribution are updated as follows:

\begin{equation}\label{betadistributionupdate}
\left\{ {\begin{array}{*{20}{c}}
{{\alpha _i} = {\alpha _i} + 1}&{{\rm the{\kern 1pt} {\kern 1pt} {\kern 1pt} {\kern 1pt} feedback{\kern 1pt} {\kern 1pt} {\kern 1pt} {\kern 1pt} is{\kern 1pt} {\kern 1pt} {\kern 1pt} {\kern 1pt} a{\kern 1pt} {\kern 1pt} {\kern 1pt} {\kern 1pt} reward}}\\
{{\beta _i} = {\beta _i} + 1}&{{\rm the{\kern 1pt} {\kern 1pt} {\kern 1pt} {\kern 1pt} feedback{\kern 1pt} {\kern 1pt} {\kern 1pt} {\kern 1pt} is{\kern 1pt} {\kern 1pt} {\kern 1pt} {\kern 1pt} a{\kern 1pt} {\kern 1pt} {\kern 1pt} {\kern 1pt} penalty}}
\end{array}\begin{array}{*{20}{c}}
{\left( {i = 1,2} \right)}
\end{array}} \right.
\end{equation}

The BLA is briefly described as follows: The agent of BLA estimates the reward probability of every action utilizing two Beta distributions, denoted as ${\bf Beta\left( {{\alpha _1},{\beta _1}} \right)}$ and ${\bf Beta\left( {{\alpha _2},{\beta _2}} \right)}$.  During the learning process, two variables are generated according to the two Beta distributions, while the action of a higher variable is selected. Then, the reward of taking the chosen action is computed to update the parameters of the two Beta distribution.

The goal of RL is to find an optimal policy that maximize the expected sum of discounted return:

\begin{equation}\label{goalofFSA}
{\pi ^*} = \mathop {\arg \max {E^\pi }}\limits_\pi  \left[ {\sum\limits_{t = 0}^{T - 1} {{\gamma ^{t + 1}}\widetilde r\left( {{{\widetilde s}_t},{{\widetilde s}_{t + 1}}} \right)} } \right].
\end{equation}

\begin{figure*}[!t]
    \normalsize
    \begin{align}\label{Probabilityofarm1}
    p_1^{s_{BLA}^n} = \frac{{\left( {\alpha _1^n + \beta _1^n - 1} \right)!\left( {\alpha _2^n + \beta _2^n - 1} \right)!}}{{\left( {\alpha _1^n - 1} \right)!\left( {\beta _1^n - 1} \right)!\left( {\alpha _1^n + \beta _1^n + \alpha _2^n + \beta _2^n - 2} \right)!}} \times \sum\limits_{j = \alpha _2^n}^{\alpha _2^n + \beta _2^n - 1} {\frac{{\left( {j + \alpha _1^n - 1} \right)!\left( {\beta _1^n + \alpha _2^n + \beta _2^n - j - 2} \right)!}}{{j!\left( {\alpha _2^n + \beta _2^n - 1 - j} \right)!}}} .
    \end{align}
    \hrulefill \vspace*{0pt}
\end{figure*}

The details of BLA based action selection can be found in Fig.~\ref{blaactionselectionleft}. The first advantage of BLA compared with other learning automata (LA) algorithms is computational simplicity, achieved by relying implicitly on Bayesian reasoning principles. The second advantage of BLA is that for the two actions case, the action selected from BLA can be proven optimal. In our scenario, we assume that the computation task is computed locally as arm 1, while the task is offloaded to the MEC server to be computed as arm2. Given the state $s_{BLA}^n = \left( {\alpha _1^n,\beta _1^n,\alpha _2^n,\beta _2^n} \right)$, the action selection probability of local computing is given in Eq.~(\ref{Probabilityofarm1}).

\begin{figure*}[!t]
\normalsize
\begin{equation}\label{Probabilityofarm2}
\begin{aligned}
{P_1}\left( {{e_1} > {e_2}} \right) &= \sum\limits_{i = 0}^{{\alpha _1} - 1} {\frac{{B\left( {{\alpha _2} + i,{\beta _1} + {\beta _2}} \right)}}{{\left( {{\beta _1} + i} \right)B\left( {1 + i,{\beta _1}} \right)B\left( {{\alpha _2},{\beta _2}} \right)}}}  \\
& = \sum\limits_{i = 0}^{{\beta _2} - 1} {\frac{{B\left( {{\beta _1} + i,{\alpha _1} + {\alpha _2}} \right)}}{{\left( {{\alpha _2} + i} \right)B\left( {1 + i,{\alpha _2}} \right)B\left( {{\alpha _1},{\beta _1}} \right)}}} \\
\end{aligned}
\end{equation}
\hrulefill \vspace*{0pt}
\end{figure*}

The details of the proposed BLA-MAQ-learning algorithm are in~{\bf Algorithm~\ref{BLA-MAQL_cacheMEC}}. However, the optimal action may not always return a higher reward than the suboptimal actions at a certain time instant when we select it. Thus, the self-correcting character of BLA is of great value. According to~\cite{Granmo2008}, the BLA is instantaneously self-correcting for two-armed bernoulli bandit problems. Here it is expressed formally as follows:

\begin{theorem}\label{theorem:instantaneously_self-correcting}
Assume that local computing is the optimal solution, and the expected value of the Beta distribution associated with offload computing, $\left( {{{\alpha _2^n} \mathord{\left/
 {\vphantom {{\alpha _2^n} {\left( {\alpha _2^n + \beta _2^n} \right)}}} \right.
 \kern-\nulldelimiterspace} {\left( {\alpha _2^n + \beta _2^n} \right)}}} \right)$ is approaching ${r_2}$, while $\left( {{{\alpha _1^n} \mathord{\left/
 {\vphantom {{\alpha _1^n} {\left( {\alpha _1^n + \beta _1^n} \right)}}} \right.
 \kern-\nulldelimiterspace} {\left( {\alpha _1^n + \beta _1^n} \right)}}} \right)$ is less than $r_2$. The BLA is capable of increasing the probability of local computing after each action selection. Moreover, BLA is instantaneously self-correcting.

\begin{IEEEproof}
See Appendix B~.
\end{IEEEproof}
\end{theorem}

\begin{remark}\label{remark:DRL}
Due to the symmetry of the two actions (i.e., local computing and offload computing), the BLA based action selection scheme is capable of reducing the probability of a non-optimal action under the situation that the probability of choosing a non-optimal action is higher than that of optimal action. In other words, if the time slot is extensively, and the total number of taken actions is infinite, then we can obtain the optimal action eventually.
\end{remark}

For further generalization, we present the following~\textbf{Theorem~\ref{theorem:the_optimal_selection}}, which describes the optimal solution of Bayesian learning automata in the two-action cases.

\begin{theorem}\label{theorem:the_optimal_selection}

In the cache-aided MEC scenario, given the two action selection between local computing and offloading computing, the BLA based offloading decision scheme is capable of converging to only choosing the optimal selection i.e., $\mathop {\lim }\limits_{n \to \infty } p_1^{s_{BLA}^n} \to 1$, if ${r_1}>{r_2}$.

\begin{IEEEproof}
See Appendix C~.
\end{IEEEproof}
\end{theorem}

\begin{remark}\label{remark:DRL}
Given the assumption that the probabilities of two actions are not equal.~\textbf{Theorem~\ref{theorem:the_optimal_selection}} asserts that the action trained from the BLA based action selection scheme is the optimal action.
\end{remark}

\begin{algorithm}[h]
\caption{The Proposed Algorithm in Cache-Aided NOMA-MEC networks}
\label{BLA-MAQL_cacheMEC}
\begin{algorithmic}[1]
    \REQUIRE
    \FOR{$i \in {{\cal N}_u}$}
        \STATE   Initialize state $s_i$, action $a_i$, reward function.
    \ENDFOR
    \STATE \textbf{Do while:} $t<T$;\
    \FOR{each episode}
    \FOR{each step}
    \FOR{each agent in ${{\cal N}_u}$}
        \STATE Select the actions according to \textbf{BLA based action selection scheme};\\
        8.1:~Initialize ${\alpha _1} = 1,{\beta _1} = 1,{\alpha _2} = 1,{\beta _2} = 1$; \\
        8.2:~Generate two values $X_1$ and $X_2$ randomly from Beta distribution ${\bf Beta}\left( {\alpha _1,\beta _1} \right)$ and ${\bf Beta}\left( {\alpha _2,\beta _2} \right)$. \\
        8.3:~If ${X_1} > {X_2}$, choose local computing, else choose offload computing. \\
        8.4:~Accept the reward from this action, then update the parameters of the Beta distribution according to Eq.~(\ref{betadistributionupdate}).
    \STATE Take the chosen action $a$, move state, then calculate reward $r$ of the action $a$ and state $s$;\
    \STATE State update: $s \leftarrow s'$;\
    \ENDFOR
    \ENDFOR
    \ENDFOR
    \STATE \textbf{return:} task offloading vector.
\end{algorithmic}
\end{algorithm}
%
%
%
\begin{table}[h]
\caption{Simulation Parameters}
	\centering
	\begin{tabular}{|c|c|c|}\hline
		Parameter&Description&Value\\\hline
		$C_{MEC}$ & computation capacity of the AP & 10 GHz/sec  \\\hline
        $N_u$ & Number of Users & 3/4  \\\hline
        $f^l$ & CPU frequency of each user & 1 GHz/sec  \\\hline
        $R_{i,j}$ & computation offloading size & 500 bits  \\\hline
        $B$ & Bandwidth & 20MHz\cite{Chen2017TWC}  \\\hline
        ${{P_t}}$ & Transmit power of the AP & 20 dBm\cite{Chen2017TWC}  \\\hline
        ${\sigma ^2}$ & Gaussian noise power & -95 dBm\cite{Chen2017TWC}  \\\hline

	\end{tabular}

	\label{table2}
\end{table}

\begin{figure} [t!]
 \centering
 \includegraphics[width=3.5in]{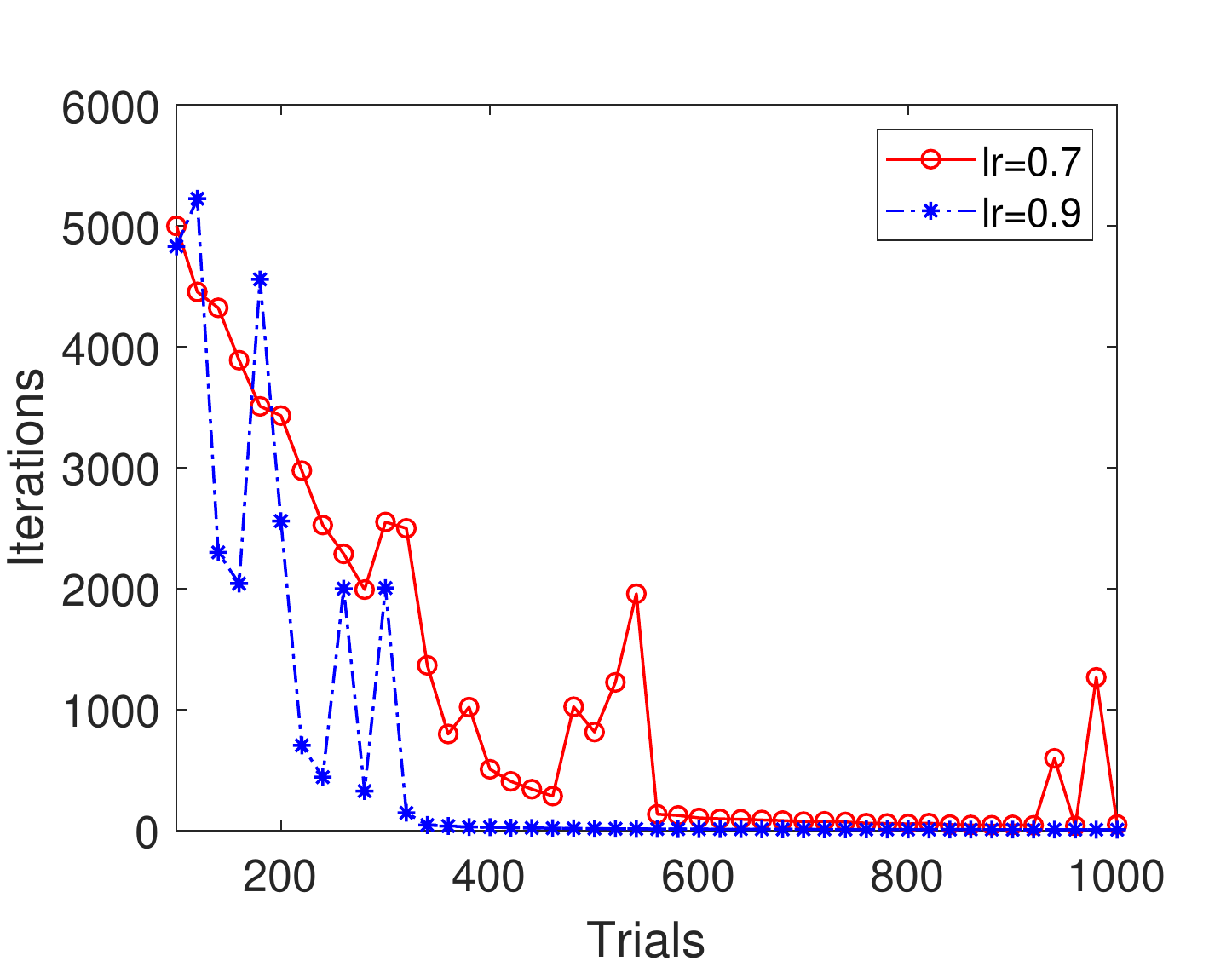}
 \centering
 \caption{Training loss of the proposed LSTM for task popularity prediction.}\label{training_loss_LSTM}
\end{figure}

\begin{figure*}

    \centering

    \subfigure[Task popularity prediction (goal equals to 0.1).]{
    \begin{minipage}{7cm}
    \centering
    \includegraphics[width=7cm]{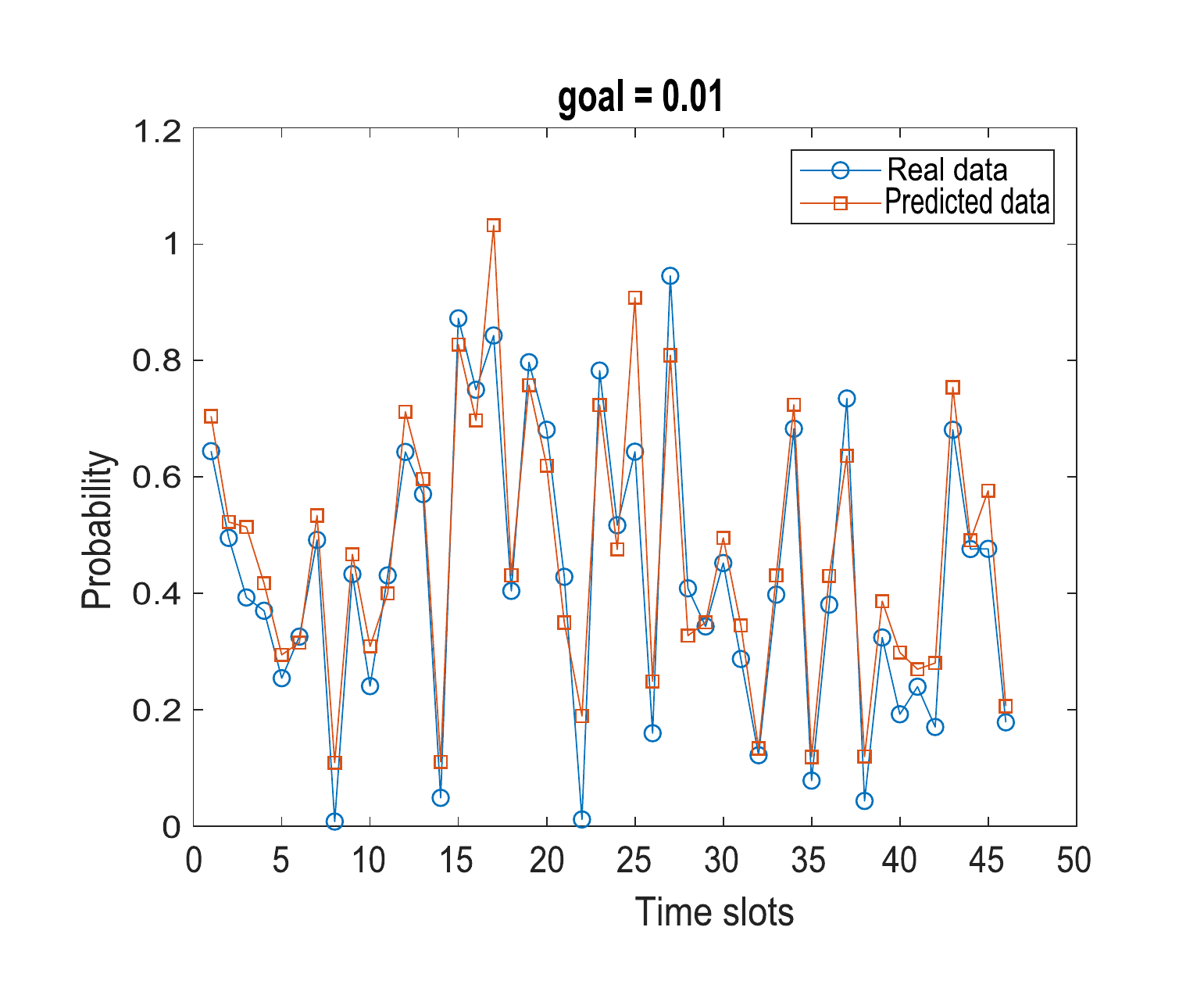}
    \end{minipage}
    }
    \subfigure[Task popularity prediction (goal equals to 0.01).]{
    \begin{minipage}{7cm}
    \centering
    \includegraphics[width=7cm]{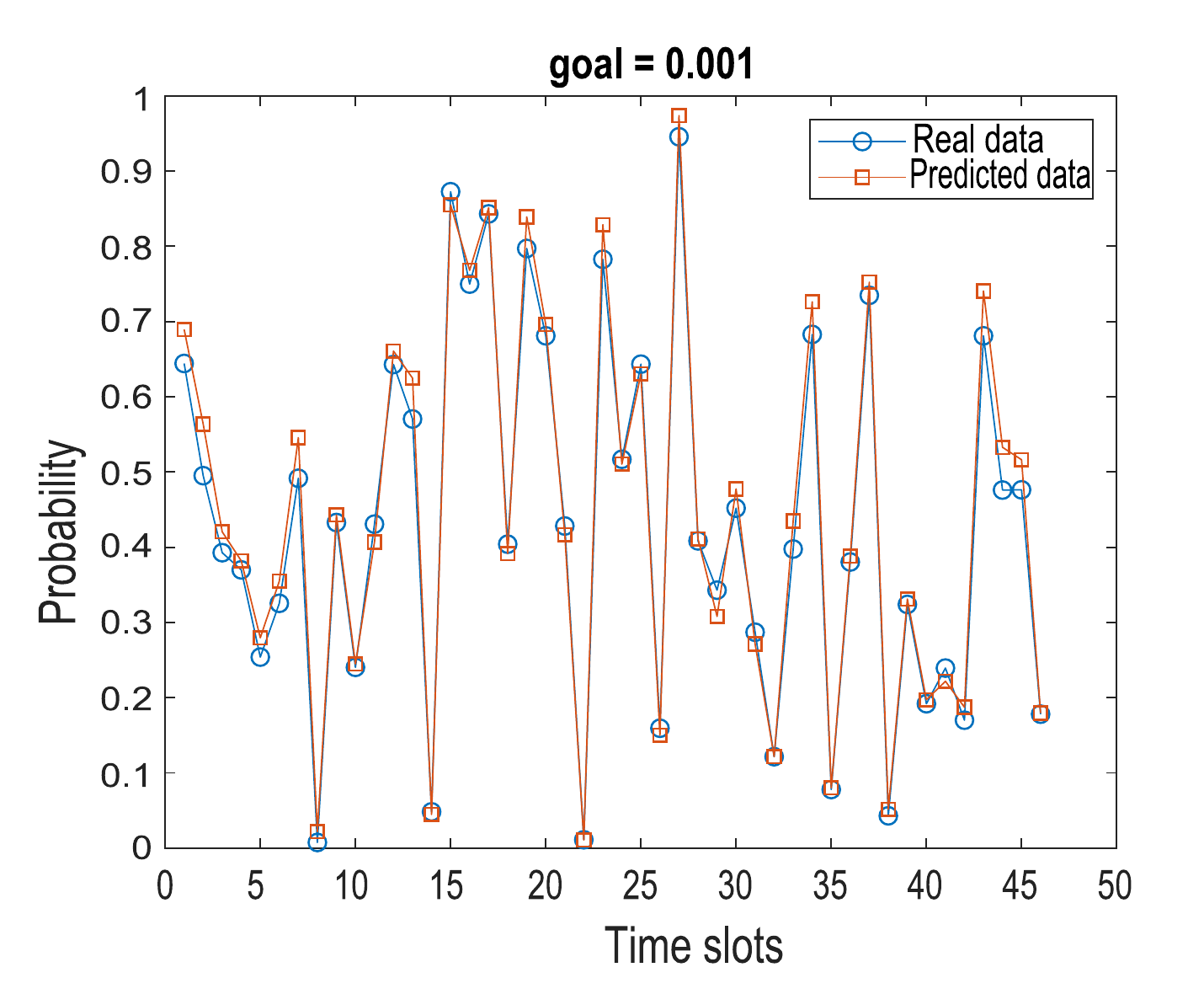}
    \end{minipage}
    }
    \caption{Simulation results of task popularity prediction using LSTMs.}\label{performanceofLSTM2}
\end{figure*}

\subsection{Complexity of the proposed algorithm}\label{subsection:Complexity}

The computational complexity analysis of the proposed BLA based MAQ-learning for the cache-aided MEC algorithm is evaluated as follows. There are $N_u$ users and $M$ computation tasks. The computational resources of the MEC server are sliced into ${N_f}$. The number of offloading decisions for $N_u$ users and caching decisions for the AP are ${2^{N_u}}$ and ${2^M}$, respectively. Thus, the complexity for MAQ-learning is $O\left( {N_u^2{2^{{N_u}}}{{N_u}^{{N_f}}} {{N_f}} } \right)$. The complexity of BLA is $O\left( {{2^{{N_u}}}} \right)$. Therefore, the complexity of the proposed BLA base MAQ-learning is $O\left( {N_u^2{2^{2{N_u}}}{{N_u}^{{N_f}}} {{N_f} } } \right)$.

%
\vspace{-0.3cm}
\section{Simulation Results}\label{section:numeralresult}

In this section, we present extensive simulation results to quantify the performance of the proposed SAQ-learning and BLA-based MAQ-learning algorithms for task offloading and computation resource allocation in a cache-aided NOMA MEC system. The simulation parameters settings are as given in Table~\ref{table2} unless otherwise stated. Due to the fact that the user's movement during requests for a computing service operates on a limited scale, we consider the situation that the positions of the users and the AP are fixed. However, different users may have different interests in different tasks. Hence, we first evaluate the proposed LSTMs for task popularity prediction. In the simulation, the positions of users in our scenario are randomly distributed within a square region with length of a side 300m. The number of users is set to 3 and 4. The AP location is in the centre of the area. The bandwidth is 20MHz. The computation capacity of the AP is $F$=10GHz/sec. Hereinafter, the size of the task input of users follows the uniform distribution with ${R_{i,j}}\left( k \right) \in \left[ {300,800} \right]$ KB. Meanwhile the number of CPU cycles per bit for the required computational resources obeys the uniform distribution with ${R_{w,j}}\left( k \right) \in \left[ {1000,1500} \right]$ cycles/bit. The CPU frequency of each user is ${f^l}$=1GHz/sec. We compare our proposed algorithm with three traditional MEC schemes: ``\emph{Full local}" means that all the tasks are computed locally in the mobile users. ``\emph{Full offloading}" means that all the tasks are offloaded to the AP for computing. ``\emph{Conventional MEC}" means that there is no caching capacity in the AP. All the simulations are performed on a desktop with an Intel Core i7 9700K 3.6 GHz CPU and 16 GB memory.

\subsection{Simulation results of task popularity prediction}\label{subsection:task_popularity_prediction}

First, we evaluates the performance of the proposed LSTMs based task prediction algorithm. In our scenario, we assume that the user requests different tasks in different time slots. Meanwhile, the tasks have different popularity. We first generate the tasks' popularity through a random walk model. Fig.~\ref{training_loss_LSTM} demonstrates the loss of the network with different learning rates. According to the figure, the proposed algorithm is capable of converging to a stable loss. The proposed LSTM converges faster with a higher learning rate. This is also justified in \textbf{Remark~\ref{remark:LSTM}}. Fig.~\ref{performanceofLSTM2} compares real task popularity and predicted popularity. According to Fig.~\ref{performanceofLSTM2}, we obtain better performance when we reduce the value of the goal.

\subsection{Simulation results of the proposed SAQ-learning algorithm for cache-aided NOMA MEC}\label{subsection:task_popularity_prediction}

\begin{figure} [t!]
 \centering
 \includegraphics[width=3.5in]{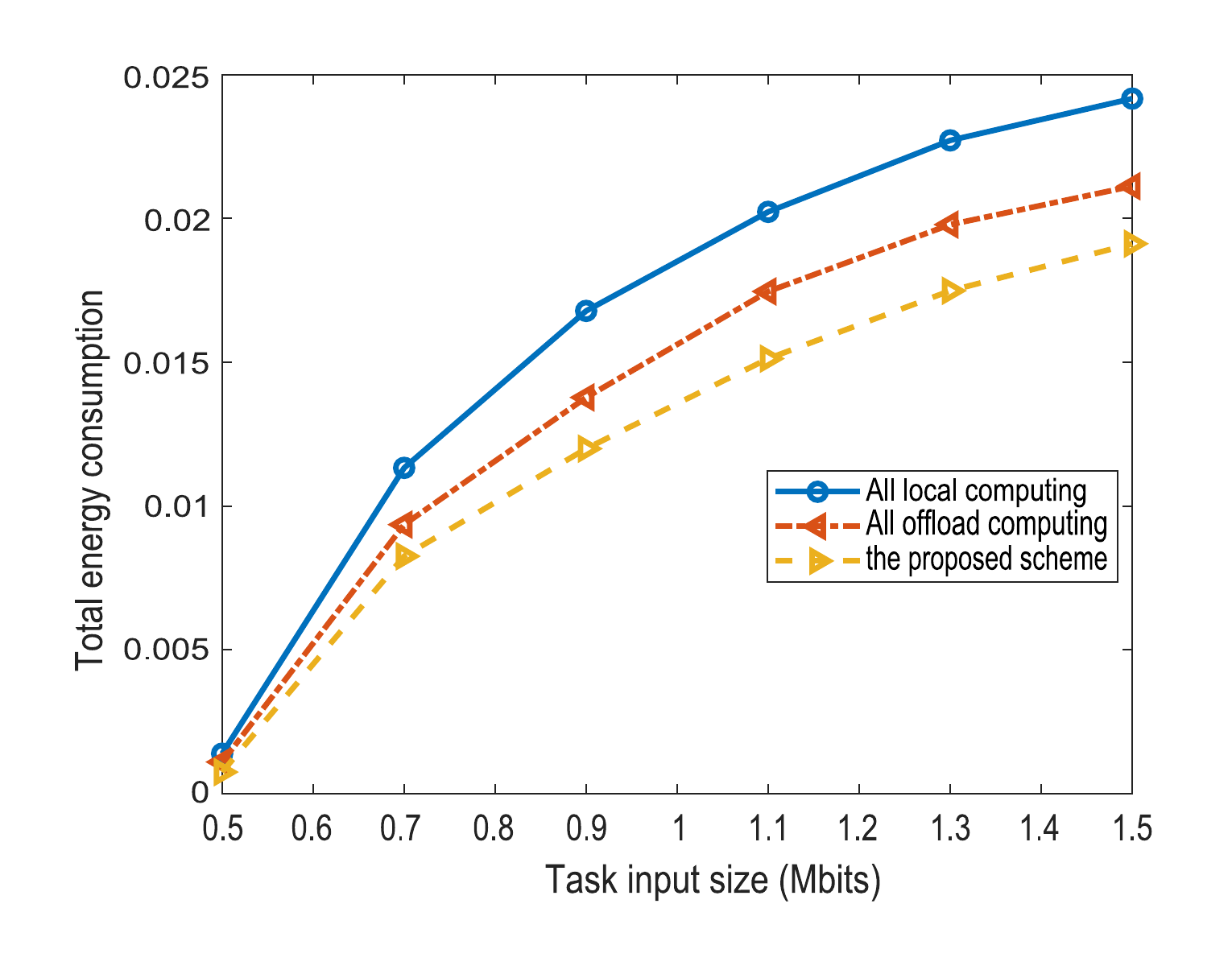}
 \centering
 \caption{Total transmit energy consumption vs. task input size.}\label{Energy_tasksizee}
\end{figure}

The simulation results of the proposed SAQL algorithm for cache-aided NOMA MEC is shown in Fig.~\ref{Energy_tasksizee} and Fig.~\ref{Energy_AP}. The total energy consumption vs. the task input size is shown in Fig.~\ref{Energy_tasksizee}. A larger task input size requires more computing energy both for the mobile users and the MEC server. This is also justified in (\ref{localcomputenergy}) and (\ref{ecl}). In Fig.~\ref{Energy_tasksizee}, the proposed cache-aided NOMA MEC outperforms both the all local computing and the all offload computing schemes. In Fig.~\ref{Energy_AP}, the proposed cache-aided NOMA MEC outperforms conventional MEC. The reason is that, in cache-aided MEC, the reusable task computing results are stored in the AP, which reduces the offloading and computing energy.
\vspace{-0.3cm}
\subsection{Simulation results of the proposed BLA-MAQ-learning algorithm for cache-aided NOMA MEC}\label{subsection:task_popularity_prediction}

\begin{figure} [t!]
 \centering
 \includegraphics[width=3.5in]{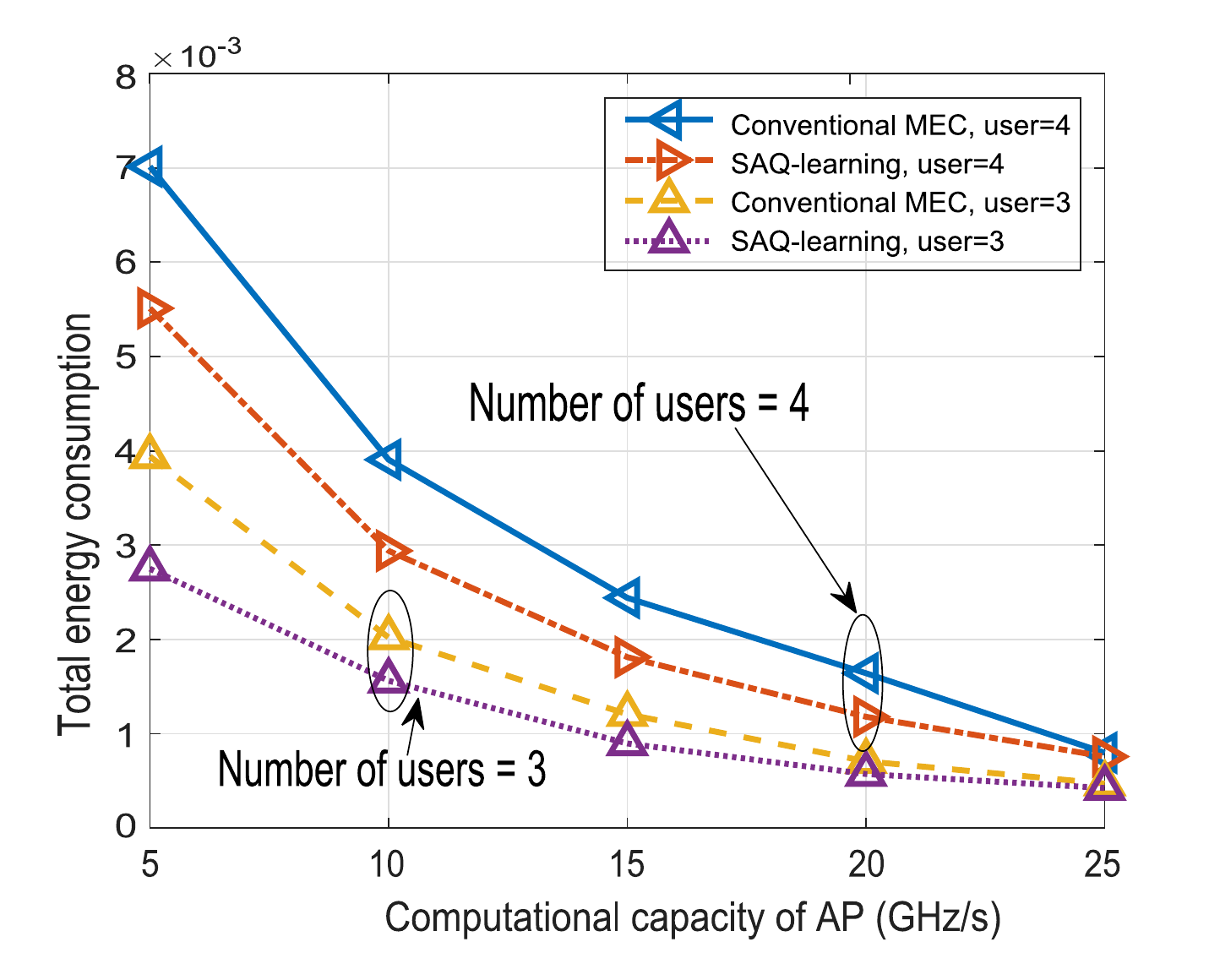}
 \centering
 \caption{Total energy consumption vs. the computation capacity of the AP.}\label{Energy_AP}
\end{figure}

\begin{figure} [t!]
 \centering
 \includegraphics[width=3.5in]{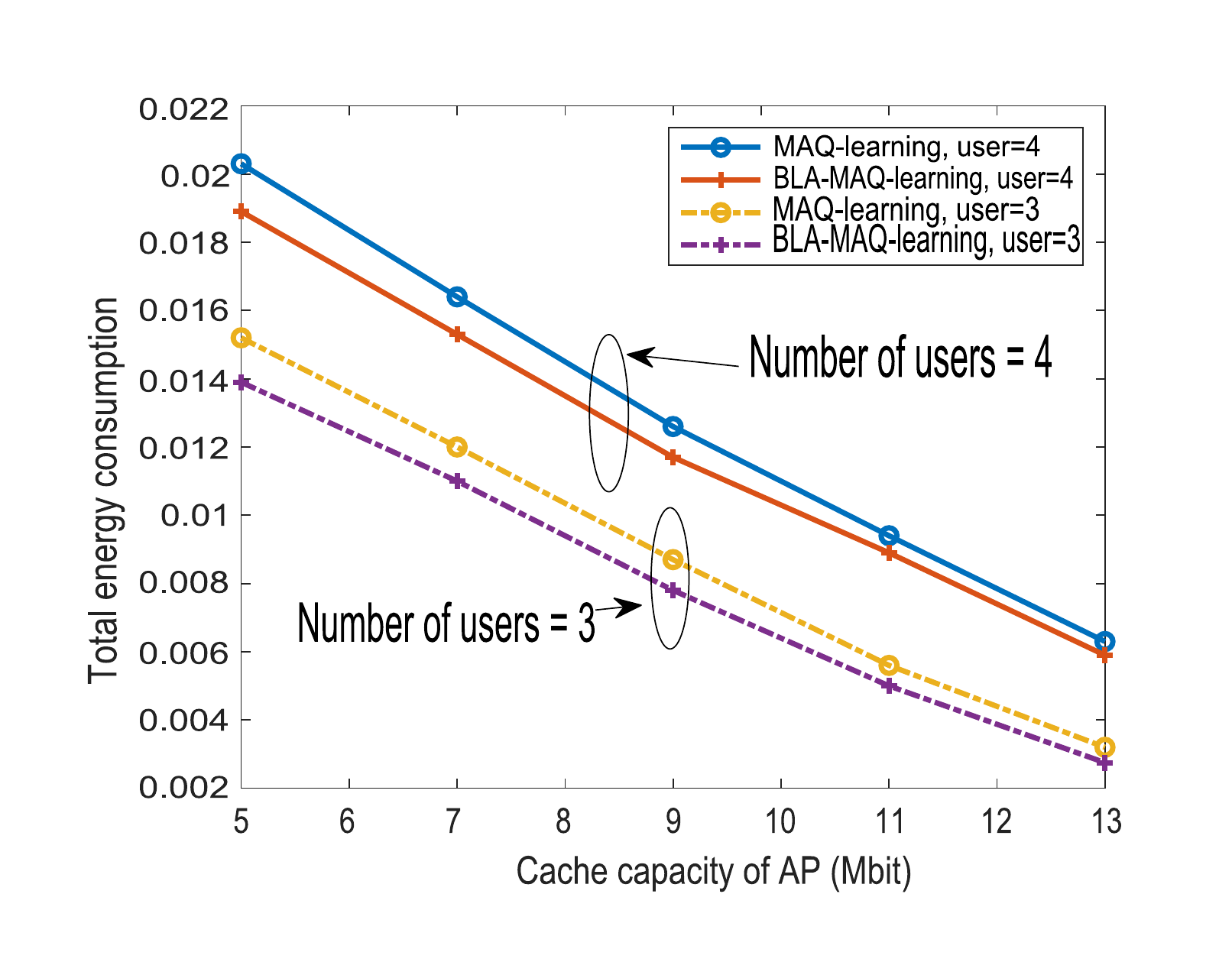}
 \centering
 \caption{Total transmit energy consumption vs. cache capacity of the AP.}\label{Energy_cache}
\end{figure}

The simulation results of the proposed BLA-MAQ-learning algorithm for cache-aided NOMA MEC is shown in Fig.~\ref{Energy_cache}. As can be seen from Fig.~\ref{Energy_cache}, the total energy consumption decreases with the increase of caching capacity. This is because more computation results are stored in the AP, moreover, the offloading consumption and computing consumption of the same tasks required by other users are all reduced. As can be seen from Fig.~\ref{Energy_AP}, the total energy consumption decreases sharply with more computational capacity. Meanwhile, compared with Fig.~\ref{Energy_cache}, the total energy consumption decrease faster than increasing the cache capacity. This shows that increasing the computational capacity of the AP is a more efficient way of reducing total energy consumption compared with increasing the cache capacity of the AP.

\begin{figure} [t!]
 \centering
 \includegraphics[width=3.5in]{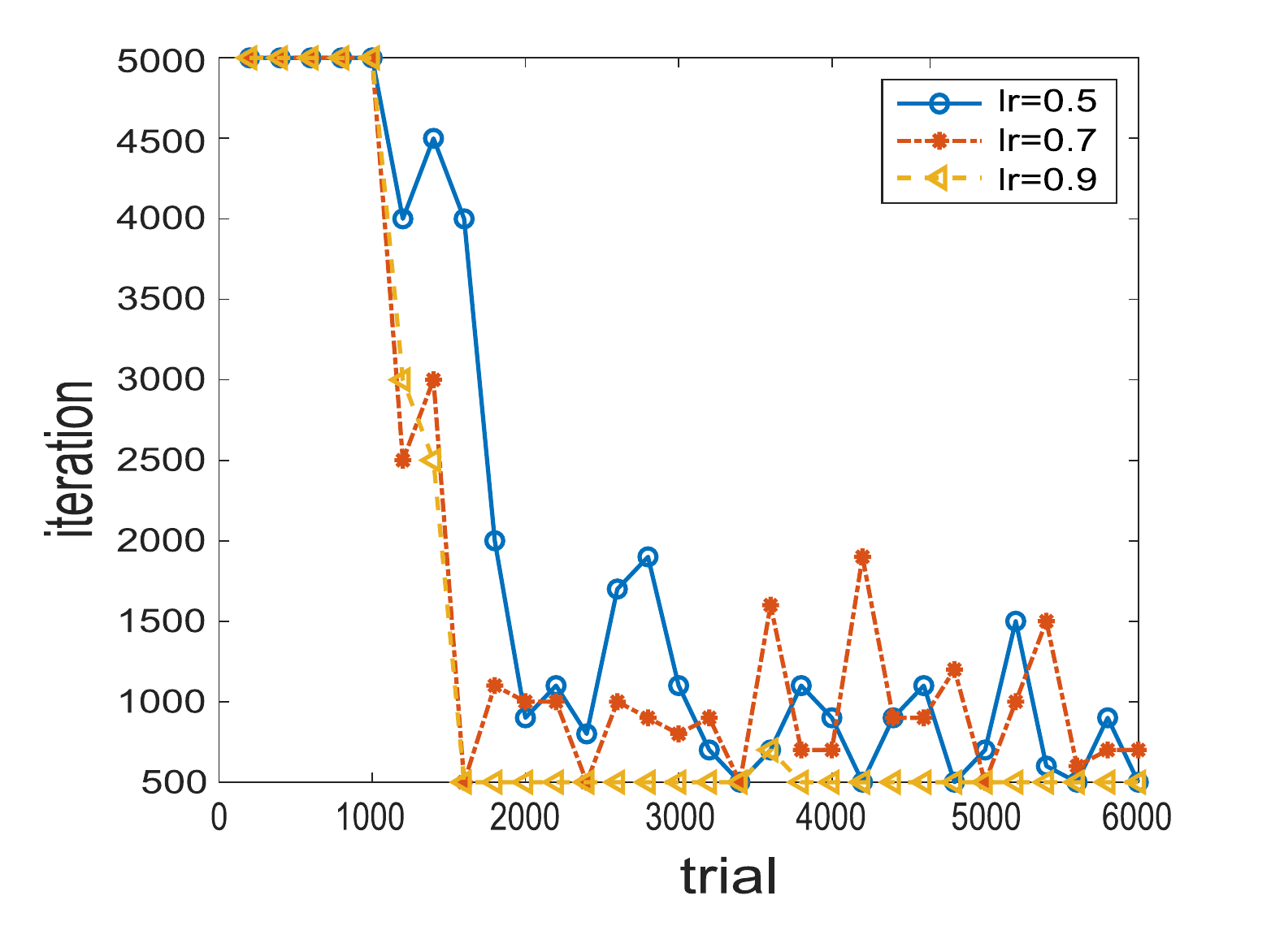}
 \centering
 \caption{The convergence of the proposed algorithm.}\label{Convenge_proposedalgorithm}
\end{figure}

Finally, the convergence of the proposed algorithm is shown in Fig.~\ref{Convenge_proposedalgorithm}. As can be seen from Fig.~\ref{Convenge_proposedalgorithm}, the proposed algorithm converges faster with a higher learning rate. However, higher learning rate may cause overfit.
\vspace{-0.3cm}
\section{Conclusion}\label{section:Conclusion}
In this paper, a cache-aided NOMA MEC framework is proposed. In order to reduce the total energy consumption in cache-aided NOMA MEC networks, a joint long-term optimization problem is formulated, subject to caching and computing resources constraints in the AP. Three parameters are optimized in the formulated MINL problem. To make task offloading decisions of users and resource allocation for the AP, a SAQ-learning based resource allocation algorithm scheme was proposed. For the task offloading problem, a BAL-MAQ-learning algorithm is proposed, in which, the BLA based action selection scheme is adopted for every agent to obtain optimal action in every state. Simulation results demonstrated the performance of the proposed framework and algorithm. Our extensive simulation results demonstrated that increasing computation capacity of the AP was a more efficient method to reduce the total energy consumption compared with increasing caching capacity. One promising extension of this work is to consider more complicated joint learning algorithms for networks with multiple MEC servers, that require cooperation between the caching resource and computing resource. Moreover, incorporating the optimization of uploading power and computing allocation can further improve energy efficiency of multi-server networks, which is another promising future research direction.
\vspace{-0.3cm}
\section*{Appendix~A: Proof of Proposition~\ref{proposition1}} \label{Appendix:A}
In Eq.~(\ref{weightvector2}), the partial derivative contains two parts: the first is the output ${\bf {{o_t}}}$, and the second part is the memory cell state ${\bf   {{C_t}} }$. The detail is in Eq.~(\ref{weightvector3}). The partial derivatives of ${{\partial \left( {{{\bf{o}}_{\bf{t}}}} \right)} \mathord{\left/
 {\vphantom {{\partial \left( {{{\bf{o}}_{\bf{t}}}} \right)} {\partial \left( {{{\bf{w}}_{\bf{C}}}} \right)}}} \right.
 \kern-\nulldelimiterspace} {\partial \left( {{{\bf{w}}_{\bf{C}}}} \right)}}$ are given in Eq.~({\ref{weightvector4}}). Based on Eq.~(\ref{outputgate1}), we can write a recursive equation as in Eq.~(\ref{weightvector5}). Combining Eq.~(\ref{forgetgate}),~(\ref{inputgate1}),~(\ref{inputgate2}), we obtain Eq.~(\ref{weightvector6})~(\ref{weightvector7})~(\ref{weightvector8}). Next, we can obtain Eq.~(\ref{weightvector2}). Finally, the proof is complete.

\begin{figure*}[!t]
\normalsize
\begin{equation}\label{weightvector3}
\begin{aligned}
\frac{{\partial \left( {{{\bf{o}}_{\bf{t}}}\tanh \left( {{{\bf{C}}_{\bf{t}}}} \right)} \right)}}{{\partial \left( {{{\bf{w}}_{\bf{C}}}} \right)}} &= \frac{{\partial \left( {{{\bf{o}}_{\bf{t}}}} \right)}}{{\partial \left( {{{\bf{w}}_{\bf{C}}}} \right)}}\tanh \left( {{{\bf{C}}_{\bf{t}}}} \right) + {{\bf{o}}_{\bf{t}}}\frac{{\partial \left( {\tanh \left( {{{\bf{C}}_{\bf{t}}}} \right)} \right)}}{{\partial \left( {{{\bf{w}}_{\bf{C}}}} \right)}}\\
& = \frac{{\partial \left( {{{\bf{o}}_{\bf{t}}}} \right)}}{{\partial \left( {{{\bf{w}}_{\bf{C}}}} \right)}}\tanh \left( {{{\bf{C}}_{\bf{t}}}} \right) + {{\bf{o}}_{\bf{t}}}\frac{{\partial \left( {\tanh \left( {{{\bf{C}}_{\bf{t}}}} \right)} \right)}}{{\partial \left( {{{\bf{C}}_{\bf{t}}}} \right)}}\frac{{\partial \left( {{{\bf{C}}_{\bf{t}}}} \right)}}{{\partial \left( {{{\bf{w}}_{\bf{C}}}} \right)}}.\\
\end{aligned}
\end{equation}
\hrulefill \vspace*{0pt}
\end{figure*}

\begin{figure*}[!t]
\normalsize
\begin{equation}\label{weightvector4}
\begin{aligned}
{\bf \frac{{\partial \left( {{o_t}} \right)}}{{\partial \left( {{w_C}} \right)}}} &= {\bf \frac{{\partial \left( {\sigma \left( {{W_o}\left[ {{h_{t - 1}},{x_t}} \right] + {b_o}} \right)} \right)}}{{\partial \left( {{w_C}} \right)}}}\\
& = {\bf \frac{{\partial \left( {\sigma \left( {{W_o}\left[ {{o_{t - 1}}*\tanh \left( {{C_{t - 1}}} \right),{x_t}} \right] + {b_o}} \right)} \right)}}{{\partial \left( {{w_C}} \right)}}}\\
& = {\bf \frac{{\partial \left( \sigma  \right)}}{{\partial \left( {{W_o}\left[ {{h_{t - 1}},{x_t}} \right] + {b_o}} \right)}}\left[ {{W_o}\frac{{\partial \left( {{o_{t - 1}}} \right)}}{{\partial \left( {{w_C}} \right)}}\tanh \left( {{C_{t - 1}}} \right) + {W_o}{o_{t - 1}}\frac{{\partial \left( {\tanh \left( {{C_{t - 1}}} \right)} \right)}}{{\partial \left( {{w_C}} \right)}}} \right]}.
\end{aligned}
\end{equation}
\hrulefill \vspace*{0pt}
\end{figure*}


\begin{figure*}[!t]
\normalsize
\begin{equation}\label{weightvector5}
{\bf \frac{{\partial \left( {{C_t}} \right)}}{{\partial \left( {{w_C}} \right)}} \\
= {C_{t - 1}}\frac{{\partial \left( {{f_t}} \right)}}{{\partial \left( {{w_C}} \right)}} + {f_t}\frac{{\partial \left( {{C_{t - 1}}} \right)}}{{\partial \left( {{w_C}} \right)}} + {d_t}\frac{{\partial \left( {{i_t}} \right)}}{{\partial \left( {{w_C}} \right)}} + {i_t}\frac{{\partial \left( {{d_t}} \right)}}{{\partial \left( {{w_C}} \right)}}.}
\end{equation}
\hrulefill \vspace*{0pt}
\end{figure*}
\begin{figure*}[!t]
\normalsize
\begin{equation}\label{weightvector6}
\begin{aligned}
{\bf \frac{{\partial \left( {{f_t}} \right)}}{{\partial \left( {{w_C}} \right)}} = \frac{{\partial \left( {{\sigma _f}} \right)}}{{\partial \left( {{W_f}\left[ {{h_{t - 1}},{x_t}} \right] + {b_f}} \right)}}\left[ {{W_f}\frac{{\partial \left( {{o_{t - 1}}} \right)}}{{\partial \left( {{w_C}} \right)}}\tanh \left( {{C_{t - 1}}} \right) + {W_f}{o_{t - 1}}\frac{{\partial \left( {\tanh \left( {{C_{t - 1}}} \right)} \right)}}{{\partial \left( {{w_C}} \right)}}} \right].}
\end{aligned}
\end{equation}
\hrulefill \vspace*{0pt}
\end{figure*}
\begin{figure*}[!t]
\normalsize
\begin{equation}\label{weightvector7}
\begin{aligned}
{\bf \frac{{\partial \left( {{i_t}} \right)}}{{\partial \left( {{w_C}} \right)}} = \frac{{\partial \left( {{\sigma _f}} \right)}}{{\partial \left( {{W_i}\left[ {{h_{t - 1}},{x_t}} \right] + {b_i}} \right)}}\left[ {{W_i}\frac{{\partial \left( {{o_{t - 1}}} \right)}}{{\partial \left( {{w_C}} \right)}}\tanh \left( {{C_{t - 1}}} \right) + {W_i}{o_{t - 1}}\frac{{\partial \left( {\tanh \left( {{C_{t - 1}}} \right)} \right)}}{{\partial \left( {{w_C}} \right)}}} \right].}
\end{aligned}
\end{equation}
\hrulefill \vspace*{0pt}
\end{figure*}
\begin{figure*}[!t]
\normalsize
\begin{equation}\label{weightvector8}
\begin{aligned}
{\bf \frac{{\partial \left( {{d_t}} \right)}}{{\partial \left( {{w_C}} \right)}} = \frac{{\partial \left( {{\sigma _f}} \right)}}{{\partial \left( {{W_d}\left[ {{h_{t - 1}},{x_t}} \right] + {b_d}} \right)}}\left[ {{W_d}\frac{{\partial \left( {{o_{t - 1}}} \right)}}{{\partial \left( {{w_C}} \right)}}\tanh \left( {{C_{t - 1}}} \right) + {W_d}{o_{t - 1}}\frac{{\partial \left( {\tanh \left( {{C_{t - 1}}} \right)} \right)}}{{\partial \left( {{w_C}} \right)}}} \right].}
\end{aligned}
\end{equation}
\hrulefill \vspace*{0pt}
\end{figure*}

\vspace{-0.3cm}
\section*{Appendix~B: Proof of Theorem~\ref{theorem:instantaneously_self-correcting}} \label{Appendix:B}

In each iteration, the computation task is either computed locally or is offloaded to the MEC server for computing. The probability that the agent chooses local computing at iteration $n$ is $p_1^{{r_2}}\left[ {\alpha _1^n,\beta _1^n} \right]$, where $\left( {\alpha _1^n,\beta _1^n} \right)$ are the parameters of the Beta distribution associated with local computing. According to the characteristic of the Beta distribution, the probability of local computing (i.e., expected value) is ${{\mathbb E}_{local}} = {{\alpha _1^n} \mathord{\left/
 {\vphantom {{\alpha _1^n} {\left( {\alpha _1^n + \beta _1^n} \right)}}} \right.
 \kern-\nulldelimiterspace} {\left( {\alpha _1^n + \beta _1^n} \right)}}$. In addition, the probability of offloaded computing is $1 - p_1^{{r_2}}\left[ {\alpha _1^n,\beta _1^n} \right]$. The feedback of local computing provides a reward with probability ${r_1}$, and a penalty with probability $1-{r_1}$. Meanwhile, if offload computing is chosen by the agent, we have $p_1^{{r_2}}\left[ {\alpha _1^{n + 1},\beta _1^{n + 1}} \right] = p_1^{{r_2}}\left[ {\alpha _1^n,\beta _1^n} \right]$, because ${r_2}$ is given, and there is no feedback from local computing. Therefore we have:

\begin{equation}\label{LAaa1}
\begin{array}{l}
{\mathbb E}\left\{ {p_1^{{r_2}}\left[ {\alpha _1^{n + 1},\beta _1^{n + 1}} \right]|p_1^{{r_2}}\left[ {\alpha _1^n,\beta _1^n} \right]} \right\}\\
 = \left( {p_1^{{r_2}}\left[ {\alpha _1^n + 1,\beta _1^n} \right]{r_1} + p_1^{{r_2}}\left[ {\alpha _1^n,\beta _1^n + 1} \right]\left( {1 - {r_1}} \right)} \right)\\
 = \left( {p_1^{{r_2}}\left[ {\alpha _1^n,\beta _1^n} \right] + p_1^{{r_2}}\left[ {\alpha _1^n,\beta _1^n} \right]\left( {1 - p_1^{{r_2}}\left[ {\alpha _1^n,\beta _1^n} \right]} \right)} \right).
\end{array}
\end{equation}

Then, by dividing the left and right side of the inequality, we can obtain Eq. (\ref{LAaa2}).


\begin{figure*}[!t]
\normalsize
\begin{align}\label{LAaa2}
\begin{array}{l}
E\left[ {p_1^{{r_2}}\left( {\alpha _1^{n + 1},\beta _1^{n + 1}} \right)|p_1^{{r_2}}\left( {\alpha _1^n,\beta _1^n} \right)} \right] > p_1^{{r_2}}\left[ {\alpha _1^n,\beta _1^n} \right]\\
 \Leftrightarrow p_1^{{r_2}}\left( {\alpha _1^n + 1,\beta _1^n} \right){r_1} + p_1^{{r_2}}\left( {\alpha _1^n,\beta _1^n + 1} \right)\left( {1 - {r_1}} \right) + \left( {1 - p_1^{{r_2}}\left( {\alpha _1^n,\beta _1^n} \right)} \right) > 1\\
 \Leftrightarrow p_1^{{r_2}}\left( {\alpha _1^n + 1,\beta _1^n} \right){r_1} + p_1^{{r_2}}\left( {\alpha _1^n,\beta _1^n + 1} \right)\left( {1 - {r_1}} \right) > p_1^{{r_2}}\left( {\alpha _1^n,\beta _1^n} \right)
\end{array}
\end{align}
\hrulefill \vspace*{0pt}
\end{figure*}

In order to prove {\bf Theorem~\ref{theorem:instantaneously_self-correcting}}, we need to prove that the following is true:
\begin{equation}\label{LAaa3}
\begin{array}{l}
{\mathbb E}\left\{ {p_1^{{r_2}}\left[ {\alpha _1^{n + 1},\beta _1^{n + 1}} \right]|p_1^{{r_2}}\left[ {\alpha _1^n,\beta _1^n} \right]} \right\} > p_1^{{r_2}}\left[ {\alpha _1^n,\beta _1^n} \right]\\
 \Leftrightarrow \frac{{\alpha _1^n}}{{\alpha _2^n + \beta _2^n}} < {r_1}.
\end{array}
\end{equation}

According to the characteristics of the Beta distribution, the probability density function (PDF) of the beta distribution is
\begin{equation}\label{LAaa4}
f\left( {x;\alpha ,\beta } \right) = \frac{{{x^{\alpha  - 1}}{{\left( {1 - x} \right)}^{\beta  - 1}}}}{{\int_0^1 {{u^{\alpha  - 1}}{{\left( {1 - u} \right)}^{\beta  - 1}}du} }} = \frac{{\Gamma \left( {\alpha  + \beta } \right)}}{{\Gamma \left( \alpha  \right)\Gamma \left( \beta  \right)}}{x^{\alpha  - 1}}{\left( {1 - x} \right)^{\beta  - 1}}.
\end{equation}

\begin{figure*}[!t]
\normalsize
\begin{align}\label{LAaa8}
\begin{array}{l}
p_1^{{r_2}}\left[ {\alpha _1^n + 1,\beta _1^n} \right]{r_1} + p_1^{{r_2}}\left[ {\alpha _1^n,\beta _1^n + 1} \right]\left( {1 - {r_1}} \right) > p_1^{{r_2}}\left[ {\alpha _1^n,\beta _1^n} \right]\\
 \Leftrightarrow \left( {1 - F\left( {{r_2};\alpha _1^n + 1,\beta _1^n} \right)} \right){r_1} + \left( {1 - F\left( {{r_2};\alpha _1^n,\beta _1^n + 1} \right)} \right)\left( {1 - {r_1}} \right) > 1 - F\left( {{r_2};\alpha _1^n,\beta _1^n} \right)\\
 \Leftrightarrow F\left( {{r_2};\alpha _1^n,\beta _1^n} \right) - {r_1}F\left( {{r_2};\alpha _1^n + 1,\beta _1^n} \right) - \left( {1 - {r_1}} \right)F\left( {{r_2};\alpha _1^n,\beta _1^n + 1} \right) > 0\\
 \Leftrightarrow F\left( {{r_2};\alpha _1^n,\beta _1^n} \right) - {r_1}\left( {F\left( {{r_2};\alpha _1^n,\beta _1^n} \right) + \frac{{\left( {\alpha _1^n + \beta _1^n - 1} \right)!}}{{\alpha _1^n!\left( {\beta _1^n - 1} \right)!}}{{\left( {1 - x} \right)}^{\beta _1^n}}{x^{\alpha _1^n}}} \right) - ...\\
\left( {1 - {r_1}} \right)\left( {F\left( {{r_2};\alpha _1^n,\beta _1^n} \right) + \frac{{\left( {\alpha _1^n + \beta _1^n - 1} \right)!}}{{\beta _1^n!\left( {\alpha _1^n - 1} \right)!}}{{\left( {1 - x} \right)}^{\beta _1^n}}{x^{\alpha _1^n}}} \right)\\
 \Leftrightarrow \frac{{\alpha _1^n}}{{\alpha _1^n + \beta _1^n}} < {r_1}.
\end{array}
\end{align}
\hrulefill \vspace*{0pt}
\end{figure*}

The expected value (mean) of a Beta distribution random variable $X$ is
\begin{equation}\label{LAaa5}
{\mathbb E}\left( X \right) = \int_o^1 {xf\left( {x;\alpha ,\beta } \right)dx}  = \int_o^1 {x\frac{{\Gamma \left( {\alpha  + \beta } \right)}}{{\Gamma \left( \alpha  \right)\Gamma \left( \beta  \right)}}{x^{\alpha  - 1}}{{\left( {1 - x} \right)}^{\beta  - 1}}dx} .
\end{equation}

According to the fact that:
\begin{equation}\label{LAaa6}
\int_o^1 {\frac{{\Gamma \left( {\alpha  + \beta  + 1} \right)}}{{\Gamma \left( {\alpha  + 1} \right)\Gamma \left( \beta  \right)}}{t^{\alpha  - 1}}{{\left( {1 - t} \right)}^{\beta  - 1}}dt}  = 1.
\end{equation}

\begin{equation}\label{LAaa6.1}
F\left( {x;\alpha  + 1,\beta } \right) = F\left( {x;\alpha ,\beta } \right) + \frac{{\left( {\alpha  + \beta  - 1} \right)!}}{{\alpha !\left( {\beta  - 1} \right)!}}{\left( {1 - x} \right)^\beta }{x^\alpha },
\end{equation}
\begin{equation}\label{LAaa6.2}
F\left( {x;\alpha ,\beta  + 1} \right) = F\left( {x;\alpha ,\beta } \right) + \frac{{\left( {\alpha  + \beta  - 1} \right)!}}{{\beta !\left( {\alpha  - 1} \right)!}}{\left( {1 - x} \right)^\beta }{x^\alpha }.
\end{equation}

The cumulative distribution function is given by:
\begin{equation}\label{LAaa7}
F\left( {x;\alpha ,\beta } \right) = \sum\limits_{k = \alpha }^{\alpha  + \beta  + 1} {\frac{{\left( {\alpha  + \beta  + 1} \right)!}}{{k!\left( {\alpha  + \beta  - 1 - k} \right)!}}{x^k}{{\left( {1 - x} \right)}^{\alpha  + \beta  - 1 - k}}} .
\end{equation}

Then, the probability ${p_1}\left( {\alpha _1^n,\beta _1^n|{r_2}} \right)$ is given by
\begin{equation}\label{LAaa9}
p_1^{{r_2}}\left( {\alpha _1^n,\beta _1^n} \right) = P\left( {X_1^N > {r_2}|\alpha _1^n,\beta _1^n} \right) = 1 - F\left( {{r_2};\alpha _1^n,\beta _1^n} \right).
\end{equation}

Therefore, substituting Eq.~(\ref{LAaa9}) in Eq.~(\ref{LAaa3}), we can get Eq. (\ref{LAaa8}). Therefore, Eq.~(\ref{LAaa3}) is correct. The BLA based action selection scheme is capable of self-correcting to the local computing under the condition that the probability of choosing offload computing is higher than that of local computing.

\vspace{-0.3cm}
\section*{Appendix~C: Proof of Theorem~\ref{theorem:the_optimal_selection}} \label{Appendix:C}

In order to prove {\bf Theorem~\ref{theorem:the_optimal_selection}}, first we need to make an assumption, which is that there is only one optimal action, i.e., $P\left( {{r_1} = {r_2}} \right) = 0$. If the probabilities of the two actions are the same, according to \textbf{Theorem~\ref{theorem:instantaneously_self-correcting}}, the probability of selecting the optimal action will not increase given the worst situation (i.e., the probability of non-optimal action is higher than that of optimal action). In other words, the self-correcting feature of BLA vanishes.

Since all the quantities involved in Eq.~(\ref{Probabilityofarm1}) are positive, we obtain $\alpha _2^n > 0$. According to \textbf{Theorem~\ref{theorem:instantaneously_self-correcting}}, the expected value ${\mathbb E}\left( {{X_i}} \right) = \frac{{\alpha _i^n}}{{\beta _i^n + \alpha _i^n}}$ approaches $r_i$ with the increase of time, i.e.:
\begin{equation}\label{LAaa11}
{\mathbb E}\left( {{X_1}} \right) = \frac{{\alpha _1^n}}{{\beta _1^n + \alpha _1^n}} = {r_1},
\end{equation}
\begin{equation}\label{LAaa12}
{\mathbb E}\left( {{X_2}} \right) = \frac{{\alpha _2^n}}{{\beta _2^n + \alpha _2^n}} = {r_2}.
\end{equation}

Combining Eq.~(\ref{LAaa11}) and Eq.~(\ref{LAaa12}), we get

\begin{equation}\label{LAaa13}
\beta _1^n = \frac{{\left( {1 - {r_1}} \right)\alpha _1^n}}{{{r_1}}}.
\end{equation}

According to~\cite{Christoffer2010}, the probability of chosen local computing is given by

\begin{equation}\label{LAaa11}
\begin{aligned}
p_1^{s_{BLA}^n}{\rm{ }} = \frac{{\left( {\beta _1^n} \right)!\left( {\alpha _2^n} \right)!}}{{\left( {\beta _1^n + \alpha _2^n} \right)!}} = \frac{{\left( {\frac{{\left( {1 - {r_1}} \right)\alpha _1^n}}{{{r_1}}}} \right)!\left( {\alpha _2^n} \right)!}}{{\left( {\frac{{\left( {1 - {r_1}} \right)\alpha _1^n}}{{{r_1}}} + \alpha _2^n} \right)!}} = 1.
\end{aligned}
\end{equation}

The proof is complete.

\vspace{-0.3cm}
\bibliographystyle{IEEEtran}
\bibliography{mybib}

\end{document}